\documentclass[
   pre,aps,
   amsfonts,
   amssymb,
   amsmath,
   superscriptaddress,
   eqsecnum,
   showpacs,
   preprintnumbers,
   byrevtex]{revtex4}
\usepackage{color}
\usepackage{graphicx}

\begin{document}

\newcommand{\vphi}{\varphi}
\newcommand{\bq}{\begin{equation}}
\newcommand{\be}{\begin{equation}}
\newcommand{\ba}{\begin{eqnarray}}
\newcommand{\eq}{\end{equation}}
\newcommand{\ee}{\end{equation}}
\newcommand{\ea}{\end{eqnarray}}
\newcommand{\tchi} {{\tilde \chi}}
\newcommand{\tA} {{\tilde A}}
\newcommand{\sech} { {\rm sech}}
\newcommand{\pstar}{\mbox{$\psi^{\ast}$}}
\newcommand {\bPsi}{{\bar \Psi}}
\newcommand {\bpsi}{{\bar \psi}}
\newcommand {\tu} {{\tilde u}}
\newcommand {\tv} {{\tilde v}}
\newcommand{\dq}{{\dot q}}
\preprint{LA-UR 12-24007} 

\title{Nonlinear Dirac equation solitary waves in external fields }
\author{Franz G.  Mertens}
\email{franzgmertens@gmail.com}
\affiliation{Physikalisches Institut, Universit\"at Bayreuth, D-95440 Bayreuth, Germany} 
\author{Niurka R. Quintero} 
\email{niurka@us.es} 
\affiliation{IMUS and Departamento de Fisica Aplicada I, E.P.S. Universidad de Sevilla, 41011 Sevilla, Spain}
\author{Fred Cooper}
\email{cooper@santafe.edu}
\affiliation{Santa Fe Institute, Santa Fe, NM 87501, USA}
\affiliation{Theoretical Division and Center for Nonlinear Studies, 
Los Alamos National Laboratory, Los Alamos, New Mexico 87545, USA}
\author{Avinash Khare} 
\email{khare@iiserpune.ac.in}
\affiliation{ Indian Institute of Science Education and Research, Pune 
411021, India}
\author{Avadh Saxena}
\email{avadh@lanl.gov}
\affiliation{Theoretical Division and Center for Nonlinear Studies, 
Los Alamos National Laboratory, Los Alamos, New Mexico 87545, USA}
\begin{abstract}
We consider the nonlinear Dirac equations (NLDE's)  in 1+1 dimension with 
scalar-scalar  self interaction 
$\frac{ g^2}{\kappa+1} ( {\bPsi} \Psi)^{\kappa+1}$ in the presence of various external electromagnetic fields. Starting from the exact solutions for the unforced problem we study the behavior of  solitary wave solutions to the NLDE in the presence of a wide variety of  fields in a variational approximation depending on collective coordinates which allows the position, width and phase of these waves to vary in time.  We find that in this approximation the position $q(t)$  of the center of the solitary wave obeys the usual behavior of a relativistic point particle in an external field.  For time independent external fields we find that the  energy of the solitary wave is conserved but not the momentum which becomes a function of time.  We postulate that similar to the nonlinear Schr{\"o}dinger equation (NLSE)  that a  sufficient dynamical condition for instability to arise is that 
$ dP(t)/d \dq(t) < 0$.  Here $P(t)$ is the momentum of the solitary wave, and $\dq$ is the velocity of the center of the wave in the collective coordinate approximation. We found for our choices of external potentials we always have $ dP(t)/d \dq(t) > 0$ so  when instabilities  do occur they are due to a different source.   We investigate the accuracy of our variational approximation using numerical simulations of the NLDE and find that when the forcing term is small and we are in a regime where the solitary wave is stable, that the behavior of the solutions of the collective coordinate equations agrees very well with the numerical simulations.  We found numerically that the time evolution of the collective coordinates of the solitary wave in our numerical simulations, namely the position of the average charge density and the momentum of the solitary wave, provide good indicators for when the solitary wave first becomes unstable.  Namely, when these variables stop being smooth functions of time ($t$) then the solitary wave starts distorting in shape. 

\end{abstract}

\maketitle

\section{Introduction}
Classical solutions of nonlinear field equations have a long history as a 
model of extended particles~\cite{ref:extended, Enz, soler1}.   In 1970, 
Soler \cite{soler1} proposed that the self-interacting 4-Fermi theory was an 
interesting model for extended fermions.  Later, Strauss and Vasquez 
\cite{ref:strauss}  were able to study the stability of this model under 
dilatation and found the domain of stability for the Soler solutions. 
Solitary waves in the 1+1 dimensional nonlinear Dirac equation (NLDE) have 
been studied \cite{ref:Lee,ref:Nogami} in the past in the case of  massive 
Gross-Neveu~\cite{ref:GN} (with $N=1$, i.e. just one localized fermion) and 
massive Thirring~\cite{ref:TM} models).  
In those studies it was found that these equations have solitary wave 
solutions for both scalar-scalar (S-S) and vector-vector (V-V) interactions. 
The interaction between solitary waves of different initial charge was 
studied in detail  for the S-S case 
in the work of Alvarez and Carreras \cite{ref:numerical} 
by Lorentz boosting the static solutions and allowing them to scatter.
Recently we extended the solutions previously found to a more general 
interaction of the form  $\frac{ g^2}{\kappa+1} ( {\bPsi} \Psi)^{\kappa+1}$
\cite{NLDE}. For the non-relativistic limit of the NLDE, namely the nonlinear 
Schr{\"o}dinger  equation (NLSE), there have been recent studies of the behavior of the 
forced NLSE.   Using a collective coordinate (CC) theory, the authors 
found \cite{qmb, mqb, Mpreprint, cooper1} that  a {\it sufficient} 
dynamical condition for instability to arise is that 
$ dp(t)/dv < 0.$ Here  $p(t)$ is the normalized canonical momentum 
$p(t) = \frac{1}{M(t)} \frac {\partial L}{\partial {\dot q} }$,  
$M(t) = \int dx \Psi^\star(x,t)  \Psi(x,t)$ is the mass and ${\dot q}(t) = v(t) $ is the  velocity of the solitary wave.  

One of the points we will investigate in the paper is 
whether this dynamical stability criterion is also valid for the NLDE.   There has been 
recent interest in the stability of NLDE with higher-order nonlinearity \cite{comech}.  Comech (private communication)  has been able to prove that for $\kappa =1$, the Vakhitov-Kolokolov  \cite{VK} criterion guarantees linear stability in the non-relativistic regime of the NLDE equation for solutions of the form (in the rest frame) 
$\Psi(x,t) = \psi(x) e^{-i \omega t}$ where $\omega$ is less than but approximately equal to the mass parameter $m$ in the Dirac equation.
He was also able to show linear instability in the same non-relativistic regime for $\kappa \geq 3$.  This is the first rigorous result for the Dirac equation, but it only applies  in the non-relativistic regime.  Here we want to understand if we can determine in the relativistic regime for what values of $\omega$ do the solitary waves  become unstable, with and without forcing terms even when they are stable in the 
non-relativistic regime.  What we find is that when the solitary waves are only metastable for the unforced problem, the critical time for the solitary wave to become unstable in the forced problem for weak forcing is similar to the critical time in the unforced problem. When the solitary wave maintains its basic shape, the CC equations give  a good description of the actual time evolution at all times.  This is true for weak ramp potentials, harmonic potentials, and  spatially periodic potentials, when $\omega > \omega_c$, and $\omega_c$ is the critical value above which the unforced solitary wave is stable. The  collective coordinates $q(t)$ and $P(t)$, the position and momentum of the solitary wave, are ``smooth" functions of $t$ for the external potentials we have chosen.  Their counterparts in the numerical simulation are the first moment of the charge density and the total momentum of the numerical solution.  When the numerical evolution of these counterparts to the collective coordinates start deviating from their CC values, this is a signal that the shape of the solitary wave is beginning to change.  This usually rapidly develops into non-smooth behavior of q(t) and P(t) in the numerical solution.  This is how we determine the onset of the instability time $t_c$ for the forced NLDE solitary wave.  Unfortunately, for the potentials we study we always obtain $dp/d{\dot{q}}>0$, which fulfills a {\it necessary} condition for stability.  Thus, this criterion does not yield a prediction of the instabilities. 

This paper is organized as follows: In Sec. II, we review the known exact solutions for the unforced NLDE  and discuss the conservation laws that govern their behavior.  In Sec. III we extend Bogolubsky's discussion \cite{ref:bogol} of the 
stability of these  solitary wave solutions to changes in the frequency $\omega$  for arbitrary  nonlinearity parameter $\kappa$.  In Sec. IV we consider the NLDE in external electromagnetic fields and obtain the covariant as well as rest frame  equations for the two components of the wave function for the solitons.    In Sec. V we introduce our variational method based on using for our  variational  wave functions  the exact wave functions for the solitary waves of the unforced problem, with the position, width parameter and phase of these solutions being promoted to  collective coordinates depending on time.  We write the relativistic equations for these collective coordinates which are similar to point particle relativistic dynamical equations. The potential the average position of the solitary wave sees is a particular average of the external potential weighted with the charge density.  In Sec. VI we postulate our stability criterion for an arbitrary external potential based on just solving the CC equations. This condition is a sufficient condition for instability.
In Sec. VII we examine and solve the collective coordinate (CC)  equations for three types of potentials--a ramp potential, a harmonic potential and a spatially periodic potential.   We also compare the solution to the CC equations to the numerical simulation of the NLDE
equation.  We state our conclusions in Sec. VIII. In the Appendix we discuss  Integral identities that are obeyed by rest frame solutions  of the unforced problem. 

\section{review of exact solutions to the  NLDE}
In this section we review the exact solutions to the NLDE, using the notation 
of \cite{NLDE}.  
We are interested in solitary wave solution of the NLDE given by 
\bq
(i \gamma^{\mu} \partial_{\mu} - m) \Psi +g^2 (\bPsi  \Psi)^{\kappa} \Psi 
= 0 \>. \label{nlde1}
\eq
These equations  can be derived in a standard fashion from the Lagrangian density
\bq
\mathcal{L} =  \left(\frac{i}{2}\right) [\bPsi \gamma^{\mu} \partial_{\mu} \Psi 
-\partial_{\mu} \bPsi \gamma^{\mu} \Psi] - m \bPsi \Psi 
+ \frac{g^2}{\kappa+1} (\bPsi \Psi)^{\kappa+1}  \>.
\eq
For solitary wave solutions,  the  field $\Psi$ goes to zero at infinity.    It is sufficient to go into 
the rest frame, since the theory is Lorentz invariant and the moving solution can be obtained by a Lorentz boost.
In the rest frame we consider solutions of the form
\bq \label{eq2.3}
\Psi(x,t) = e^{-i\omega t} \psi(x) .
\eq
We are interested in bound state solutions that correspond to positive frequency $\omega \geq 0$ and which have energies in the rest frame less than the mass parameter $m$, i.e. $\omega < m$. 
In our previous paper \cite{NLDE}, we chose the representation  $\gamma_0 = \sigma_3$, $i \gamma_1= \sigma_1$. Here, to make contact with the numerical simulations paper of Alvarez and Carreras \cite{ref:numerical} we choose instead
$\gamma^0 = \sigma_3$;  ~~~$\gamma^1= i \sigma_2.$
Defining $A,B$ via:
\ba  \label{defAB}
\psi(x) &&  =  \left(  \begin{array} {cc}
      A(x) \\
      i ~B(x) \\ 
   \end{array} \right) =
R(x) \left(\begin{array}{c}\cos \theta \\ i \sin \theta \end{array}\right),
\ea
we obtain the following equations for $A$ and $B$.
\ba
&& \frac{dA}{dx} + (m+\omega ) B - g^2(A^2-B^2)^{ \kappa} B=0\,, \nonumber \\
&&\frac{dB}{dx} + (m-\omega ) A - g^2(A^2-B^2)^{ \kappa} A=0\,. \nonumber \\
\ea
A first integral of these equations can be obtained from energy-momentum conservation. The energy momentum tensor  is given by 

\bq  \label{emc}
 T^{\mu \nu} = \frac{i}{2} \left[ \bPsi \gamma^\mu \partial^ \nu \Psi  -  \partial^\nu \bPsi  \gamma^\mu  \Psi \right]  - g^{\mu \nu} \cal{L}.
 \eq
Energy-momentum conservation follows from the equations of motion and we have that 
\bq
\partial_\mu T^{\mu \nu} =0.
\eq
The energy density is given by 
\bq
 {\cal H} =T^{00} = -\frac{i}{2} \left[  \bPsi  \gamma^1 \partial_x \Psi-\partial_x \bPsi \gamma^1 \Psi \right]+ m \bPsi \Psi - \mathcal{L}_I  \equiv  h_1+ h_2- h_3 , \label{eq:hdensity}
\eq
where 
\bq
 \mathcal{L}_I  =   \frac{g^2}{ \kappa+1} (\bpsi \psi)^{ \kappa+1} . \ 
\eq
For our rest frame solutions we have that $T^{00}$ is independent of time. Therefore the equation
\bq
\partial_{t} T^{00} + \partial_{x} T^{10}= 0,
\eq
leads to the result
\bq
 T^{10} = const.
\eq
For our rest frame solution  $T^{01} =  \frac{i}{2} \left[ \bPsi \gamma^0 \partial^x \Psi  -  \partial^x \bPsi  \gamma^0 \Psi \right] $ is  also independent of time so that 
\bq
\partial_{t} T^{01} + \partial_{x} T^{11}= 0, 
\eq 
leads to the result 
\bq
 T^{11} = const. 
\eq
Now using (\ref{eq2.3}) we obtain 
\bq
T^{11} = \omega  \psi^\dag \psi - m  \bpsi \psi + \mathcal{L}_I; ~~  \mathcal{L}_I  =   \frac{g^2}{ \kappa+1} (\bpsi \psi)^{ \kappa+1} . \ 
\label{eq:t112}
\eq
For solitary wave solutions vanishing at infinity the constants $T^{10}$ and $T^{11}$ are zero and we get the useful first integral:
\bq
T^{11}= \omega  \psi^\dag \psi - m  \bpsi \psi + \mathcal{L}_I = 0. \label{eq:t11}
\eq
Multiplying the  equation of motion  on the left by $\bPsi$   and using (\ref{eq2.3}) 
we have that:
\bq
( \kappa+1)  \mathcal{L}_I = - \omega  \psi^\dag \psi + m  \bpsi \psi  - \bpsi i \gamma^1 \partial_1 \psi .  \label{eq:motion}
\eq
Therefore we can rewrite $T^{11} = 0$ as 
\bq
\omega   \kappa  \psi^\dag \psi - m  \kappa    \bpsi \psi  - \bpsi i \gamma^1 \partial_1 \psi = 0.
\eq

From Eqs. (\ref{eq:t11}) and (\ref{eq:motion}) one has the relationship:
\bq
  \kappa  \mathcal{L}_I =   -\bpsi i \gamma^1 \partial_x \psi  .  \label{eq:rel}
\eq
From this we have
\bq
h_3 = \frac{1} { \kappa}  h_1   \label{eq:relation}
\eq
and in particular for $ \kappa=1$, $ {\cal H} = m \bpsi \psi$.

 In terms of $R, \theta$
one has 
\bq
 \bpsi i \gamma_1 \partial_1 \psi = \psi^\dag \psi \frac{d \theta} {dx} . 
 \eq
 This leads to the simple differential equation for $\theta$  for solitary waves 
 \bq
 \frac{d \theta} {dx} = -  \omega _ \kappa+ m_ \kappa \cos 2 \theta  ;    
  ~~\omega _ \kappa  \equiv  \kappa~ \omega    ; ~~ m_ \kappa =  \kappa~m . 
 \eq
The solution is (in this section and what follows we will choose the position of the solitary wave to be initially at $x_0=0.$) 
\bq
\theta(x) =  \tan^{-1} ( \alpha \tanh \beta_{ \kappa} x) , 
\eq
where 
\bq
\alpha = \left( \frac{m_ \kappa - \omega _ \kappa}{m_ \kappa + \omega _ \kappa} \right) ^{1/2}
=  \left( \frac{m - \omega }{m + \omega } \right) ^{1/2}, ~~ \beta_{ \kappa} = (m_ \kappa^2 - \omega _ \kappa^2)^{1/2} . 
\eq
Thus we have
\ba
\tan \theta(x)  &&=  \alpha \tanh \beta_{ \kappa} x , \nonumber \\
\sin^2 \theta(x) &&  = \frac{(m-\omega) \sinh^2 
\beta_\kappa x}{m \cosh 2 \beta_{\kappa} x + \omega} ; ~~
\cos^2 \theta(x) =  \frac{(m+\omega) \cosh^2 \beta_{\kappa} x}{m \cosh 2 \beta_{\kappa} x + \omega} ,  \label{theta} 
\ea
where we have used the identities:
\ba
1+ \alpha^2 \tanh^2 \beta_{k} x & = & \left(\frac {m \cosh 2 \beta_{k} x + \omega }{m+\omega } \right) \sech^2\beta_{k} x  \>,
\nonumber \\
1- \alpha^2 \tanh^2 \beta_{k} x &  = & \left(\frac {\omega \cosh 2 \beta_{k} x + m}{m+\omega } \right)  \sech^2\beta_{k} x \>.
  \label{identities2}
\ea
Solving  Eq. (\ref{eq:t11}) for $R^2$ we obtain
\bq
R^2 =\left[  \frac{( \kappa+1) (m \cos 2 \theta -\omega ) }{g^2 (\cos 2 \theta)^{ \kappa+1} }\right] ^{1/ \kappa} . 
\eq
Now we have
\bq
\frac{d \theta}{dx} = \frac{\beta_{ \kappa}^2}{\omega _ \kappa+m_ \kappa \cosh 2\beta_{ \kappa} x} = -\omega _ \kappa + m_ \kappa \cos 2 \theta ,  \label{now}
\eq
where $\beta_{ \kappa} = \sqrt{m_ \kappa^2 -\omega _ \kappa^2} =  \kappa  \sqrt{m^2-\omega ^2}$, so that 

\bq  \cos 2 \theta =  \frac{m_ \kappa+\omega _ \kappa \cosh 2 \beta_{ \kappa} x}{\omega _ \kappa+m_ \kappa \cosh 2 \beta_{ \kappa} x }=  \frac{m+\omega  \cosh 2 \beta_{ \kappa} x}{\omega +m \cosh 2 \beta_{ \kappa} x } . 
\eq
We can rewrite $R^2$ using the RHS of  Eq. (\ref{now}) as 
\bq
R^2 = \left( \frac {\omega +m \cosh 2 \beta_{ \kappa} x}{ m+\omega  \cosh 2 \beta_{ \kappa} x} \right)    \left[ \frac {( \kappa+1) \beta_{ \kappa}^2} 
{g^2  \kappa^2 (m+\omega  \cosh 2 \beta_{ \kappa} x)} \right]^{1/ \kappa} . 
\eq
Using the identities  in  Eq. (\ref{identities2}), we obtain the alternative expression
\bq
R^2 =\left( \frac  {1+\alpha^2 \tanh^2\beta_{ \kappa} x }  {1-\alpha^2 \tanh^2\beta_{ \kappa} x } \right) 
  \left[ \frac{{\rm sech}^2\beta_{ \kappa} x  ( \kappa+1) \beta_{ \kappa}^2}{g^2  \kappa^2 (m+\omega) ( 1-\alpha^2 \tanh^2\beta_{ \kappa} x )} \right ]^{1/ \kappa} . \label{eq:Rsq}
\eq
In particular for $\kappa=1$
\bq
R^2= \frac{2 (m-\omega)}{g^2}  \frac{(1+\alpha^2 \tanh^2 \beta x)}{(1-\alpha^2 \tanh^2 \beta x)^2} \sech^2 \beta x
\eq
and 
\ba \label{AB}
A^2&&= R^2 \cos^2 \theta =  \frac{2}{g^2} \frac{(m^2-\omega^2) (m+\omega) \cosh^2 \beta x}{(m+\omega \cosh 2 \beta x)^2} , \nonumber \\
B^2&&= R^2 \sin^2 \theta =  \frac{2} {g^2} \frac{(m^2-\omega^2) (m-\omega) \sinh^2 \beta x}{(m+\omega \cosh 2 \beta x)^2} . 
\ea

For arbitrary $\kappa$ we have
   \ba
A && =  \sqrt{ \frac{(m+\omega)  \cosh ^2(\kappa \beta x)}{m+\omega \cosh(2\kappa \beta x)}} 
\bigg [\frac{(\kappa+1) \beta ^2}{g^2 (m+\omega  \cosh (2\kappa \beta x))} 
\bigg ]^{\frac{1}{2\kappa}}  , \nonumber   \\ 
B && =   \sqrt{ \frac{(m-\omega) \sinh^2(\kappa \beta x)}{m+\omega \cosh(2\kappa \beta x)}} 
\bigg [\frac{(\kappa +1) \beta ^2}{g^2 (m+\omega  \cosh (2\kappa \beta x))} 
\bigg ]^{\frac{1}{2\kappa}}  . 
   \ea
  Because of Lorentz invariance we can find the solution in a frame moving with velocity $v$ with respect to the rest frame.
  The Lorentz boost  is given in terms of  the rapidity variable $ \eta$ as follows  (here $c=1$): 
  \bq
  v = \tanh \eta;~~   \gamma = \frac{1}{\sqrt{1-v^2}} = \cosh \eta; ~~ \sinh \eta =  \frac{v}{\sqrt{1-v^2}} . 
  \eq
  
  In the moving frame, the transformation law for spinors tells us that:
  \bq
  \Psi(x,t) =    \left(\begin{array}{cc}
        \cosh(\eta/2) & \sinh(\eta/2) \\
        \sinh(\eta/2) &  \cosh(\eta/2\\
     \end{array} \right)  \left(  \begin{array} {cc}
      \Psi_1^0[\gamma(x-vt), \gamma(t-vx)] \\
  \Psi_2^0[\gamma(x-vt), \gamma(t-vx)]\\ 
   \end{array} \right) , 
\eq
since
\bq
\cosh (\eta/2) = \sqrt{(1+\gamma)/2};~~  \sinh (\eta/2) = \sqrt{(\gamma-1)/2} . 
 \eq
 This in component form:
 \ba
 &&\Psi_1(x,t) = \left( \cosh(\eta/2) A(x') + i \sinh(\eta/2) B(x') \right) e^{-i\omega t'}  , \nonumber \\
&&\Psi_2 (x,t) = \left( \sinh(\eta/2) A(x') + i \cosh(\eta/2) B(x') \right) e^{-i\omega t'}  ,
\ea
where
\bq
x' = \gamma(x-vt); ~~ t' = \gamma(t-vx) . 
\eq
Note that  $\cosh^2(\eta/2)+\sinh^2(\eta/2) = \cosh \eta = \gamma$. 
\subsection{Conservation Laws of the NLDE}
The Lagrangian is invariant under the transformation of phase 
$\Psi \rightarrow e^{i \Lambda} \Psi$, which by Noether's theorem leads to the conserved current:
\bq \label{cc}
\partial_\mu j^\mu(x) = 0;  ~~~ j^\mu = \bPsi \gamma^\mu \Psi .
\eq
This leads to charge conservation:
\bq
Q = \int dx \Psi^\dag \Psi ,
\eq
which for the solitary wave solution leads to 
\bq\label{q}
Q=\int dx  (A^2+B^2) = \frac{1}{\kappa \beta }
\left[\frac{(\kappa +1)\beta^2}{g^2(m+\omega)} \right]^{1/\kappa}
I_{\kappa} (\alpha^2)\,,
\eq 
where
\ba
&&I_{\kappa}(\alpha^2)=\int^{1}_{-1} dy \frac{1+\alpha^2 y^2}{(1-y^2)^{( \kappa-1)/ \kappa}
[1-\alpha^2 y^2]^{( \kappa+1)/ \kappa}}\,  \nonumber \\
&&= B\left(\frac{1}{2},\frac{1}{\kappa}\right)
\phantom{a}_2F_1\left(1+\frac{1}{\kappa},\frac{1}{2},\frac{1}{2}+\frac{1}{\kappa};\alpha^2\right)
+\alpha^2 B\left(\frac{3}{2},\frac{1}{\kappa}\right)
\phantom{a}_2F_1\left(1+\frac{1}{\kappa},\frac{3}{2},\frac{3}{2}+\frac{1}{\kappa};\alpha^2\right) , 
\ea
and $_2F_1$ is a hypergeometric function and $B(x,k)$ denotes the beta function. 

We also have energy-momentum conservation Eq. (\ref{emc})
leading to conservation of energy and momentum:
\bq \label{pc}
E = \int T^{00} dx ; ~~ P = \int T^{01} dx . 
\eq
Because of Lorentz invariance it is sufficient to calculate the energy-momentum tensor in the comoving frame $v=0$. The energy momentum tensor in an arbitrary frame is then given by
\bq \label{tmunub}
T^{\mu \nu} = \Lambda^{\mu}_\alpha  \Lambda^{\nu}_\beta T^{\alpha \beta} ; ~~ \Lambda^{\mu}_\alpha=  
\left( \begin{array}{cc} 
      \cosh \eta & \sinh \eta \\
      \sinh \eta  & \cosh \eta \\
   \end{array} \right) . 
   \eq
   In the rest frame of the solitary wave, for the unperturbed system
 one has that 
   \bq
   T^{00} = h_1\left(1-\frac{1}{\kappa}\right)+h_2\,,
\eq
where
\bq
h_1=R^2(x)\frac{d\theta}{dx}
= \frac{\kappa \beta^2}{m+\omega\cosh(2\kappa \beta x)}
\left[\frac{(\kappa +1) \beta ^2}{g^2 (m+\omega  \cosh (2\kappa \beta x))} 
\right]^{1/\kappa}\,,
\eq
\bq
h_2=m \bpsi \psi = m(A^2-B^2) 
= m\left[\frac{(\kappa+1) \beta ^2}{g^2 (m+\omega  \cosh (2\kappa \beta x))} 
\right]^{1/\kappa}\,.
   \eq
Integrating in the rest frame, we get for the rest frame energy
\bq
E_0=H_1\left(1-\frac{1}{\kappa}\right)+H_2\,,
\eq
where
\ba\label{h_1}
&&H_1=\int dx h_1 =\frac{\beta}{m+\omega} 
\left[\frac{(\kappa+1)\beta^2}{g^2(m+\omega)} \right]^{1/\kappa} \nonumber \\
&&\times B\left(\frac{1}{2},1+\frac{1}{\kappa}\right)   
\phantom{a}_2F_1\left(1+\frac{1}{\kappa},\frac{1}{2},\frac{3}{2}+\frac{1}{\kappa};\alpha^2\right) ,
\ea
\ba\label{h_2}
&&H_2=\int dx h_2 =\frac{1}{\kappa \beta} 
\left[\frac{(\kappa +1)\beta^2}{g^2(m+\omega)} \right]^{1/\kappa} \nonumber \\
&&\times B\left(\frac{1}{2},\frac{1}{\kappa}\right)   
\phantom{a}_2F_1\left(\frac{1}{\kappa},\frac{1}{2},\frac{1}{2}+\frac{1}{\kappa};\alpha^2\right) . 
\ea
Since in the rest frame  for stationary solutions  $T^{11}=T^{01}=0$, the energy of the solitary wave in the moving frame is just
 \bq
 E= E_0 \cosh \eta = \gamma E_0; ~~ P = E_0 \sinh \eta , 
 \eq
 so that the norm $E^2 - P^2 = E_0^2=M_0^2$. 
 
 In particular, for $\kappa=1$ {\it and $m=1$}, we have that
\bq  \label{mzero}
 M_0 = \frac {2} {g^2 Q}  \sinh ^{-1}  \frac {g^2 Q} {2}; ~~Q = \frac {2  \sqrt { 1-\omega^2}} {g^2 \omega} .
  \end{equation}
We also have
\ba  \label{h12}
H_1&&  = -\frac{2 \left(\sqrt{1-\omega^2}-2 \tanh
   ^{-1}\left(\sqrt{\frac{1-\omega}{\omega+1}}\right)\right)}{g^2} , 
 \nonumber \\
H_2&&= \frac{4 \tanh ^{-1}\left(\sqrt{\frac{1-\omega }{\omega+1}}\right)}{g^2} . 
\ea
The conservation of energy-momentum implies for  the rest-frame soliton solution $\psi(x)$ certain relationships between spatial integrals of various combinations of powers of $A$ and $B$ which we will derive in the Appendix and which will be useful in simplifying our variational approach to the forced Dirac equation. 

\section{Stability of Exact Solutions}
\subsection{Stability to changes in the frequency at fixed charge}
Bogolubsky \cite{ref:bogol} suggested that the stability could be ascertained by looking at variations of the wave function,  keeping the charge fixed and seeing if the solution was a minimum (stable to that variation) or maximum (unstable to that variation)  of the Hamiltonian as a function of the parameter $\omega$. This principle has been very useful in the past to determining the stability of scalar wave equations that are Hamiltonian dynamical systems.  If the variation decreases the energy it turned out that the solitary waves were unstable.  Since in higher dimensions there are many degrees of freedom for perturbing the system, this criterion is a sufficient condition for instability.
For the Dirac case we have found from our numerical simulations that this criterion does not determine the critical $\omega$ except when $\kappa=1$ \cite{Niurka}, the case originally studied by Bogolubsky \cite{ref:bogol}.  Assuming we  know the wave function at  the value of $\omega$ corresponding to a fixed charge $Q$,   if we change the parametric dependence on $\omega$ this also changes the charge.
This can be corrected by assuming that the new wave function has a new normalization that corrects for this.
That is if we parametrize a rest frame solitary wave solution of the NLDE  which has a charge $Q[\omega]$ by
\bq
\psi_s(x,t) = \chi_s(x,\omega) e^{-i\omega t},
\eq
then we choose our slightly changed wave function to be
\ba
\tilde{\psi}[ x,t, \omega', \omega ] &&= \frac{\sqrt{Q[\omega]}}{\sqrt{ Q[\omega']}}  \chi_s(x,\omega') e^{-i\omega' t}   \equiv f(\omega',\omega)  \chi_s(x,\omega') e^{-i\omega' t}  .
\ea
Then the wave function $\tilde{\psi} [x,t,\omega',\omega] $ has the same charge as $\psi[x,t, \omega]$.  Inserting this wave function into the Hamiltonian
we get a new probe Hamiltonian $H_p$  depending on both $\omega', \omega$.  As a function of $\omega'$ this new Hamiltonian is stationary as a function of $\omega'$  at the 
value $\omega'=\omega$.  The criterion Bogolubsky proposed \cite{ref:bogol} is that the solitary wave is stable (unstable) with respect to this variation in $\omega$  according to whether this new
Hamiltonian has a minimum (maximum) at $\omega' = \omega$.   What we will find for $ \kappa=1$ is that there is a critical value of $\omega$ 
(determined by the coupling g and Q) below which the solitary wave is unstable, and this result is borne out by numerical simulations which we will present below. However, we will present in another paper numerical simulations at arbitrary $\kappa$ which suggest that this approach does not give results that coincide with the domain of stability of solutions of the unforced problem \cite{Niurka}. 
The probe Hamiltonian has the form:
\bq
H_p[\omega', \omega] = H_1[\omega'] \left( f(\omega',\omega)^2 - \frac{1}{ \kappa}  f(\omega',\omega)^{2( \kappa+1)} \right) + H_2[\omega'] f(\omega',\omega) ^2 . 
\eq

For $ \kappa=1$ we have that 
$ 
f(\omega',\omega) ^2 = \frac{\beta[\omega] \omega' }{\beta[\omega'] \omega} , 
$
where $\beta[\omega] = \sqrt{1-\omega^2}$.
We then find that the first derivative of $H_p$ with respect to $\omega'$ evaluated at $\omega' = \omega$ is indeed zero.
The second derivative evaluated at $\omega' = \omega$ leads to the following expression:
\bq
\frac{\partial^2H_p}{\partial\omega'^2}\Big|_{\omega=\omega'} = -\frac{2 \left(\sqrt{1-\omega ^2} \left(\omega ^2-3\right)+4 \tanh
   ^{-1}\left(\sqrt{\frac{1-\omega }{\omega +1}}\right)\right)}{g^2 \omega ^2 \left(\omega
   ^2-1\right)^2} . 
   \eq
   This function is zero at $\omega_c=0.697586$ and the second derivative is negative below this value of $\omega$ showing an instability.
  In our numerical simulations of the unforced NLDE \cite{Niurka}, we find that below this value the solitary waves are metastable, with the time for
  the instability to set in  increasing  exponentially  as a function of  $\omega$ for $\omega < \omega_c$. 
  
\section{NLDE in external electromagnetic fields}
We add electromagnetic interactions through the gauge covariant derivative 
\bq
i \partial_\mu \Psi  \rightarrow  ( i \partial_\mu-eA_\mu) \Psi\,,
\eq
then under the combined transformations
\bq  \Psi \rightarrow e^{i \Lambda(x,t)} \Psi; ~~ A_\mu \rightarrow A_\mu- \frac{1}{e} \partial_\mu \Lambda, ~~~ \bPsi \rightarrow \bPsi e^{-i \Lambda(x,t)}
\eq 
the Lagrangian is invariant.
Again the conserved current is given by Eq. (\ref{cc}).
  The  gauge invariant Lagrangian for the external field problem is
 \bq
L =  \left(\frac{i}{2}\right) [\bPsi \gamma^{\mu} \partial_{\mu} \Psi 
-\partial_{\mu} \bPsi \gamma^{\mu} \Psi] - m \bPsi \Psi 
+ \frac{g^2}{\kappa+1} (\bPsi \Psi)^{\kappa+1} - e \bPsi \gamma^\mu A_\mu \Psi  \>. \label{laggi}
\eq

Although energy is conserved if one has a potential that is time independent, momentum is not in the presence of explicitly spatially dependent external electromagnetic potentials. The energy-momentum tensor is again given by the relationship:
\bq  \label{emt}
 T^{\mu \nu} = \frac{i}{2} \left[ \bPsi \gamma^\mu \partial^ \nu \Psi  -  \partial^\nu \bPsi  \gamma^\mu  \Psi \right]  - g^{\mu \nu} \cal{L}.
 \eq
We can obtain the equation for the energy-momentum tensor in the presence of an external vector potential by considering the 
Dirac equation and its conjugate. 
 The NLDE in an external vector potential is given by
  \bq
 i \gamma^\mu \partial_\mu \Psi - m \Psi + g^2 (\bar \Psi \Psi)^\kappa \Psi - e \gamma^{\mu} A_{\mu} \Psi = 0,  \label{nldev}
 \eq
whereas the adjoint NLDE is
 \bq
- i \partial_\mu \bar{\Psi} \gamma^\mu - m \bar{\Psi} +g^2 (\bar \Psi \Psi)^\kappa \bar{\Psi} - e \bar{\Psi}  \gamma^{\mu} A_{\mu} = 0.  \label{anldev}
 \eq
 Multiplying Eq. (\ref{nldev}) to the left by $\bar{\Psi}_{\nu}$ and Eq. (\ref{anldev}) to the right by $\Psi_{\nu}$, and then adding  both expressions,  we obtain
 \bq
\partial_{\mu} T^{\mu \nu} =  e \bar{\Psi} \gamma^{\mu} \partial^{\nu} A_{\mu} \Psi. \label{c3}
\eq

This leads to 
 
\bq
\partial_{t} T^{01} + \partial_{x} T^{11}= - e \bar{\Psi} \gamma^{\mu} \partial_{x} A_{\mu} \Psi. \label{c1}
\eq 

Using the freedom of gauge transformation, one can choose in an arbitrary frame,  the axial gauge $A^1=0$, $eA_0 = e \phi(x) =  V(x)$. This is equivalent to the Lorentz gauge $\partial_\mu A^\mu=0$  in 1+1 dimensions. (In another frame, to determine the form of $A^\mu$  one needs to use the fact that $A^\mu$ transforms as a Lorentz vector). Integrating Eq. ({\ref{c1})  over all space, and assuming
$T^{11} (+ \infty) -  T^{11} (- \infty) = 0$ we get the force law: 
\bq
\frac {dP}{dt} = -e \int_{-\infty}^{\infty}  dx \psi^\dag \psi \frac {d \phi}{dx} = - e \left\langle  \frac {d \phi}{dx} \right\rangle , ~~ P =\int_{-\infty}^{\infty}  dx T^{01} .
\eq
Note that for a rest frame solution $T^{01}$ is independent of time.  This means for the rest frame solution $\langle \frac{dV(x)}{dx}  \rangle =0$.  This will be true for example if $V(x)$ is an even function of $x$. 
For the energy density we get
\bq
\partial_{t} T^{00} + \partial_{x} T^{10}=  e \bar{\Psi} \gamma^{\mu} \partial_{t} A_{\mu} \Psi. \label{c2}
\eq
For our potential, the r.h.s of (\ref{c2}) is again zero. This is important because  for the stationary solution, $T^{10}$ and $T^{00}$ do not depend on time. When the r.h.s. of (\ref{c2}) is  zero, $T^{10}=$ constant, with the constant being zero for a solitary wave. Therefore without lost of generality one can assume again that for rest frame solitons 
$\Psi(x) = e^{-i \omega t} \psi(x)$,
 $\psi=(A; iB)^{T}$.  This ensures that  $T^{10}=0$  for rest frame solitons. 
 In the rest frame with our choice of gauge  the Dirac equation becomes
 \bq
 i \gamma^\mu \partial_\mu \Psi - m \Psi + g^2 (\bar \Psi \Psi)^\kappa \Psi - \gamma^0 V(x) \Psi = 0.  \label{diraceq1}
 \eq
We have that $A$ and $B$ obey
\ba
&&\partial_x A + (m +\omega) B - g^2[A^2-B^2]^{\kappa} B -V(x) B =0, \label{eq1} \\ 
&&\partial_x B + (m-\omega) A - g^2[A^2-B^2]^{\kappa} A + V(x) A =0. \label{eq2}
\ea
We note that these equations are invariant under reflection $x \rightarrow -x$ provided $A$ (or $ B$) is odd and $V$ and $B $ (or $A$)  are even functions of $x$. 
In a future paper we will discuss how to obtain numerically the solitary wave solutions of Eqs. \eqref{eq1} and \eqref{eq2} for certain potentials.  

It is interesting to see how the equations for $R$ and $\theta$ as well as the energy-momentum conservation equations are modified in the presence of the external potential $V(x)$.
Setting  $A= R \cos \theta$, $B= R \sin \theta$, and multiplying (\ref{eq1}) by $\cos(\theta)$, 
(\ref{eq2}) by $\sin(\theta)$ and then adding both resulting equations, we obtain that $R(x)$ satisfies
\bq
R_{x}=-  R [m - g^2 R^{2 \kappa} \cos^{\kappa}(2 \theta)] \sin(2 \theta), \label{eq3}
\eq
i.e., the equation of $R$ is not affected directly by the potential. 
Now multiplying (\ref{eq1}) by $-\sin(\theta)$,  
(\ref{eq2}) by $\cos(\theta)$ and then adding both resulting equations, we obtain that $\theta(x)$ satisfies
\bq
\theta_{x}=\omega-[m - g^2 R^{2 \kappa} \cos^{\kappa}(2 \theta)]\cos(2 \theta)-V(x), \label{eq4}
\eq
i.e., the potential $V(x)$ explicitly appears in this equation. For $V(x)=\mu \cos(2 \theta)$,   
\bq
\theta_{x}=\omega-[m+\mu- g^2 R^{2 \kappa} \cos^{\kappa}(2 \theta)] \cos(2 \theta).
\eq



Note that for a rest frame soliton,  $T^{01}$ is independent of time, so that in the axial gauge we have from Eq. (\ref{c1})
\bq
\omega \frac {d}{dx}  [ \psi^\dag \psi] = m \frac {d}{dx}  [ \bpsi \psi]  - g^2 \frac {d}{dx}  [ \bpsi \psi] ^{\kappa+1} /(\kappa+1)  + e \phi(x) \frac {d}{dx}  [ \psi^\dag \psi] . 
\eq
So now integrating over $x$ from $ - \infty$ to $y $  we obtain the  equation
\bq
T^{11} = [\omega- e\phi(y) ]  \psi^\dag(y)  \psi(y)  - m \bpsi(y)  \psi(y)  + g^2 [ \bpsi(y)  \psi(y) ] ^{\kappa+1} /(\kappa+1) = -e \int_{- \infty}^y ~~\psi^\dag (x) \psi (x)  \frac {d \phi}{dx} 
\eq
instead of  the external force free case which yielded Eq. (\ref{eq:t11}).

   \section{Variational Ansatz for the NLDE in external Fields}
 The  gauge invariant Lagrangian for the external field problem is given by Eq. (\ref{laggi}).
Using  the freedom of gauge invariance, one can choose the axial gauge $A^1=0$,
  $eA_0 = V(x)$.
Our ansatz for the trial variational wave function is to assume that because of the smallness of the perturbation the main modification to our exact solutions to the NLDE equation  without an external field is that the parameters describing the position, momentum, boost and phase become time dependent. That is, we replace
\bq
vt \rightarrow q(t);~~ \eta \rightarrow \eta(t); ~~  \gamma  \omega v \rightarrow p(t);  ~~~  \omega t'=  \gamma\omega( t - vx) \rightarrow \phi(t) -p(t)(x-q(t)) , 
\eq
where $\phi(t) = \omega \gamma t - p(t)q(t)$.

Thus our trial wave function in component form is given by: 

\ba
&&
\Psi_1(x,t) = \left( \cosh{\frac{\eta}{2}} A(x') 
+ i \sinh{\frac{\eta}{2}} B(x') \right) e^{-i \phi + i p(x-q)} , \nonumber \\
&&\Psi_2(x,t) = \left( \sinh{\frac{\eta}{2}} A(x') 
+ i \cosh{\frac{\eta}{2}} B(x') \right) e^{-i \phi + i p(x-q)} , 
\ea
where $x' = \cosh \eta(t)~( x-q(t))$. 
Using this trial wave function we can determine the effective Lagrangian for the variational parameters.
Writing the Lagrangian density as 
\bq
\mathcal{L} = \mathcal{L}_1 + \mathcal{L}_2+\mathcal{L}_3 , 
\eq
where 
\ba \mathcal{L}_1&&= \frac{i}{2} \left( \bPsi \gamma^\mu \partial_\mu \Psi - \partial_\mu \bPsi \gamma^\mu \Psi \right)  , \nonumber \\
{\cal L}_2 &&= - m \bPsi \Psi + \frac{g^2}{\kappa+1} (\bPsi \Psi)^{\kappa+1};~~ {\cal L}_3 = -e A_0 \bPsi \gamma^0 \Psi \equiv  - V(x) \Psi^\dag \Psi . 
\ea
Integrating over $x$ and changing integration variables to $z = (x-q) \cosh \eta$ one obtains
\bq
L_1 = \int dx  \mathcal{L}_1 = Q \left( p {\dot q}+ {\dot \phi} - p \tanh \eta \right) - I_0 \left ( \cosh \eta -{\dot q} \sinh \eta \right) ,
\eq
where 
\bq
Q= \int dz[A^2(z)+B^2(z)]\,  
\eq
is as given by Eq. (\ref{q}).  Note that 
\bq
I_0 = \int dz \left( B' A-A'B  \right) = {H_1} , \label{izero}
\eq
where $H_1$ is the rest frame kinetic energy and is given by Eq. (\ref{h_1}). Here $ B'(x') = \frac{ dB(x')}{dx'}$, and  

\bq 
L_2 = \int{\cal L}_2 dx = -\frac{m}{\cosh \eta} I_1 
+ \frac{g^2}{(\kappa+1) \cosh \eta} I_2 ,
\eq
where 
\ba
I_1&& = \int dz \left( A^2(z)-B^2(z)  \right) ;~~I_2=\int dz \left( A^2(z)-B^2(z) \right)^{(\kappa+1)} \label{i2} , 
\ea
and 
\bq
L_3 = -  \int dz \rho(z)  V\left[\frac{z}{\cosh \eta} + q(t)\right] =  - U[\eta(t), q(t)]  \label{udeff} .
\eq
Putting these terms together we obtain:
\ba
&&
L = Q(p {\dot q} + {\dot \phi} - p \tanh \eta) - I_0 \left( \cosh\eta - {\dot q} \sinh \eta \right) \nonumber \\
&& -\frac{m}{\cosh \eta} I_1 +\frac{g^2}{(\kappa+1) \cosh \eta} I_2 -U[\eta(t), q(t)] . 
\ea
We now get the following Lagrange's equations:
\bq
\frac{d}{dt} \frac {\delta L} {\delta {\dot \phi}} = 0 \rightarrow 
\frac{dQ}{dt} = 0 \rightarrow  Q= {\text const}\,,
\eq
i.e. the charge is canonically conjugated to the phase $\phi$. The canonical
solitary wave momentum, which is conjugated to the solitary wave position, is

\ba
P_q &&=  \frac {\delta L} {\delta {\dot q}} =Qp + I_0 \sinh \eta,~~  \nonumber \\
\frac{dP_q}{dt}&& = Q {\dot p} + I_0 \cosh \eta\, {\dot \eta} =  \frac {\delta L} {\delta {q}} = - \frac{\partial U}{\partial q} .
\nonumber
\ea
From 
\bq 
\frac {\delta L}{\delta p} = 0 \rightarrow {\dot q} = \tanh \eta,  
\eq
which implies $\sinh \eta = \gamma {\dot q}$ and $\cosh \eta = \gamma = (1-{\dot q}^2)^{-1}$.  Also, 

\ba
 \frac {\delta L}{\delta \eta}&& = 0 \rightarrow -Q p~ \sech^2 \eta 
- I_0(\sinh \eta - {\dot q} \cosh \eta ) \nonumber \\
 && + \tanh \eta \sech \eta \bigg(mI_1-\frac{g^2}{(\kappa+1)} I_2\bigg) - \frac {\partial U}{\partial \eta} =0.
 \ea
Changing variables to $ {\dot q} = \tanh \eta$ and using
$
\frac { d {\dot q}}{d\eta} = \sech^2 \eta,
$
we obtain
\bq
Q p(t) = \gamma{\dot q}  \left(mI_1-\frac{g^2}{(\kappa+1)} I_2\right) - \frac{\partial U}{\partial {\dot q}}  \label{p1}
\eq
and
\bq
Q {\dot p} +I_0  \gamma {\dot q} =  \frac {\delta L} {\delta {q}} = - \frac{\partial U}{\partial q} . \label{dotp2}
\eq
From Eq. (\ref{p1}) we also have 
\bq
Q {\dot p}  = \frac{d (\gamma {\dot q})}{dt}  \left(mI_1-\frac{g^2}{(\kappa+1)} I_2\right)
 -  \frac {d }{dt} \frac{\partial U}{\partial {\dot q}}. \label{dotp1}
\eq
Combining Eqs.  (\ref{dotp2}), (\ref{dotp1}) we obtain an equation for the generalized force $F_{eff}$
\bq
\mu \frac{ d (\gamma {\dot q})}{dt}  = F_{eff} [q,{\dot q} ] , \label{force}
\eq
where 
\bq
\mu = mI_1-\frac{g^2}{(\kappa+1)} I_2 + I_0;~~  F_{eff} [q,{\dot q} ] 
= \frac {d }{dt} \frac{\partial U}{\partial {\dot q}} 
- \frac{\partial U}{\partial q} . \label{Feff}
\eq
Now for the NLDE without the presence of external forces, the solitary wave in the frame with $v=0$ obeys the relationship 
\cite{ref:Lee}  
\bq
\omega \psi^\dag \psi - m \bpsi \psi + \frac{g^2}{\kappa+1}
(\bpsi \psi)^{\kappa+1} =0.
\eq
For our problem this converts into 
\bq
\omega(A^2+B^2) -m(A^2-B^2) + \frac{g^2}{(\kappa+1)} (A^2-B^2)^{(\kappa+1)} =0.
\eq
Integrating this relationship we obtain:
\bq
mI_1 - \frac{g^2}{(\kappa+1)} I_2 = \omega Q ,  \label{relation1}
\eq
thus we can write Eq. (\ref{force}) as 
\bq
 \frac{ d (M {\dot q})}{dt}  = F_{eff} [q,{\dot q} ] , \label{force1}
\eq
where 
\bq
M = ( Q \omega+ I_0) \gamma = M_0 \gamma . \label{relation2}
\eq

Here $F_{eff}$ is given by Eq. (\ref{Feff}), where 
\bq
U({\dot q}, q)=  \int_{-\infty}^{\infty} dz~  V\left(q+ \frac{z}{\gamma} \right) [A^2(z)+ B^2(z)]  ,
\eq
and $\gamma = 1/\sqrt{1- {\dot q}^2}$.
We can rewrite the relativistic force equation as
\bq
\gamma^3 M_0 \ddot{q} = F_{eff} [q,{\dot q} ] .  \label{force2}
\eq
Using the rest frame identities  of the Appendix we have that
\bq
M_0 = I_0 + \omega Q .
\eq
Here $I_0$ is given by Eq. (\ref{izero}), and Q is given by Eq. (\ref{q}).

It is useful to rewrite the equation for the canonical momentum $P_q$ using the definition of $M_0$ and Eq.  (\ref{p1}) as follows:
\bq
P_q = Qp + I_0 \gamma {\dot q} = M_0 \gamma {\dot q} -  \frac{\partial U[q,{\dot q}] }{\partial {\dot q}} . \label{pq}
\eq

\subsection{Energy-momentum tensor}
The fact that the external potential  is explicitly  independent of time means that the energy of the solitary wave is independent of time.
The energy density is given by
\bq
T^{00} = \frac{i}{2} (\bpsi \gamma^0 \partial_t \psi - \partial_t \bpsi \gamma^0 \psi) -  {\cal L} . 
\eq
Straightforward integration leads to 
\bq
E = \int dx T^{00} = Q p \dq + \gamma I_0 + \frac{m}{\gamma} I_1 
- \frac {g^2}{(\kappa+1) \gamma} I_2 + U[q,\dq] . 
\eq
Using the identities  Eq. (\ref{relation1})  and Eq.(\ref{relation2}) we can rewrite this 
as
\bq
E = M_0 \gamma + Q p \dq - \gamma \omega Q \dq^2 + U[q,\dq] . 
\eq
From (\ref{p1}) and (\ref{relation1}) we have that 
\bq
Qp = \gamma \dq \omega Q - \frac{ \partial U} {\partial \dq} ,
\eq
thus we can write the energy of the solitary wave in the convenient form:
\bq
E = M_0 \gamma  + U[q, \dq] - \dq  \frac{ \partial U} {\partial \dq} . \label{econs}
\eq
The conservation of energy will be important to test our numerical integration schemes in Sec. VII.

For time independent external forces the total momentum of the solitary wave is not conserved but changes depending on the external force.  We have that 
\bq
P= \int T^{01} dx = - \frac {i}{2} \int dx ( \psi^\dag \partial_x \psi -  \partial_x \psi^\dag  \psi) .  \label{totalP}
\eq
Explicitly we obtain
\bq
P =  \gamma \dq I_0+ p Q , 
\eq
where $I_0$ is given by Eq. (\ref{izero}).
Using Eq. (\ref{p1}),  we can rewrite this as 
\bq
P = \gamma M_0 \dq - \frac {\partial U[q,\dq]}{\partial \dq} , \label{canp}
\eq
which we recognize as identical to the canonical momentum  $P_q = \frac{\delta L}{\delta \dq}$ given by Eq. (\ref{pq}).
The Lagrange equation for $P_q$ is 
\bq
{\dot P_q} = -\frac{\partial U[q,{\dot q}] }{\partial { q}}  = \frac{d}{dt} \left(M_0 \gamma {\dot q} -  \frac{\partial U[q,{\dot q}] }{\partial {\dot q}} \right).
\label{canonicalP}
\eq

\section{Stability conjecture}
For the NLSE under the influence of external forces, one could determine fairly accurately  the domains of stability of solitary wave solutions \emph{without} solving the exact partial differential equations for the given external force, but instead studying the behavior of the collective variables in a variational approximation using known solutions of the unforced problem as the trial wave functions. 
In \cite{NLDE,mqb}, it was demonstrated  that a reliable dynamical stability 
criterion for the breakup of the solitary wave under external forces was that the solitary wave will be stable if
\bq
\frac{\partial p(t) }{\partial {\dot q}(t)}  > 0.
\eq
Here $p(t)$ is the normalized momentum of the solitary wave $P(t)/M(t)$, where $M= \int dx \Psi^\star(x,t) \Psi(x,t)$ is the ``mass"  of the
solitary wave.  For the NLDE $Q$ takes the place of $M$. However, $Q$ is a conserved variable so one can use the canonical momentum $P(t)$ instead of $P( q, \dq )/Q$ to study stability. 
Using our collective coordinate theory, this leads to  the criterion that a {\it necessary} (but not sufficient) condition for stability of the solitary waves of the CC theory is that 
\bq\label{pdot}
\frac{\partial P(q,\dq)}{\partial {\dot q}}  = \gamma^3 M_0 - \frac{\partial^2 U}{\partial \dq^2} >0.
\eq

Note that the r.h.s. of Eq. (\ref{pdot}) plays the role of a time dependent mass.
The {\it sufficient}   condition for the solitary wave solution to the CC equation to be {\it unstable} in our simulations is that
\bq\label{pdotunstable}
\frac{\partial P(q,\dq)}{\partial {\dot q}}  < 0.
\eq


Following Comech's reasoning \cite{comech}, we expect that in the non-relativistic regime where $\omega$ is close to $m$ that
this criterion will be valid in determining stability in the case of external sources.  However for the NLDE we are instead studying the effect of external potentials on solitary wave motion which is quite a different problem.  For the  external potentials we have studied in this paper   Eq. (\ref{pdotunstable}) was never satisfied, so that the instabilities that we see are instead often related to the $\omega$ instabilities already present in the problem without external potentials or some other cause. Thus our hope of obtaining a simple way of determining the domain of instabilities using Eq. (\ref{pdotunstable}) was not borne out.  

\section{Simple external Potentials}
 \subsection{Simulations}

The numerical simulations have been performed by means of a
4th order Runge-Kutta method. We choose $N+1$ points starting at $n=0$
and vanishing boundary conditions $\Psi(\pm L,t)=0$.
The other parameters related with the 
discretization of the system are 
$x \in [-100,100]$, $\Delta x=0.02$, $\Delta t=0.0001$.   
For our initial conditions on the solitary wave, we use the  exact 1-solitary wave solutions of the 
unforced nonlinear Dirac equation discussed in Sec. II.  Since we would like to compare the exact numerical solution with 
the solution of the CC equations, we need to define how we determine the position of the solitary wave. In our 
numerical computation of $q(t) $ we have used the first moment of the charge, i.e. 

\bq
q(t) =Q_{1}(t)/Q(t) ,
\eq
 where 
 \bq
 Q_{1}=\int dx ~x ~\Psi^{\dag} (x)\Psi (x), ~~ Q=\int dx ~\Psi^{\dag} (x)\Psi (x).  
 \eq
For the collective variable $P(t)$ we use the definition found in Eq. (\ref{totalP}).  What we will find from our numerical simulations is that the shape of the solitary wave starts deforming once one or both of these collective variables $q(t)$ and $P(t)$ become unsmooth or rapidly varying functions of time.  Simultaneous to that happening these variables start to differ from their counterparts found solving  the CC equations.
\subsection{Linear potential (ramp potential) }
Consider the constant external force with scalar potential  $V(x) = - V_1 x$, and $V_1>0.$ 
We then have from Eq. (\ref{udeff}) that
\bq
U = - V_1  q(t)  Q ,
\eq
the force law then becomes
\bq
\frac{d}{dt} (M_0 \gamma \dq ) = V_1 Q . 
\eq
Integrating once  [starting at an initial velocity $ \dq(0)$] one has 
\bq
\frac {\dq}{\sqrt{1-\dq ^2}} = c_1 t +c_2 ,
\eq
where $c_1 = V_1 Q/M_0$ and $c_2=\frac {\dq(0)}{\sqrt{1-\dq(0) ^2}}$.  Integrating we obtain
\bq
q(t) = \frac{\sqrt{( { c_1}
   t+ { c_2})^2+1}}{ { c_1}}-\frac{\sqrt{ { c_2}^2+1}}{ { c_1}}+q(0) .
\eq

This is the standard result for a relativistic point  particle undergoing constant acceleration. 
If we choose $ q(0) = \dq( 0) = 0$, we get the simpler expression
\bq
q(t) = \frac{1}{c_1} \left[\sqrt{1+c_1^2 t^2} -1 \right]  . \label{qt}
\eq
The energy of the solitary wave is just
\bq
E = M_0 \gamma + U = M_0 , 
\eq
and the force law is now
\bq
{\dot P} = - \frac{\partial U}{\partial q}= V_1 Q , 
\eq
so that 
\bq
P = V_1 Q t . 
\eq
Since $U$ in this case is independent of $\dq$, we find from Eq.  (\ref {pdot}) that
\bq
\frac{\partial P}{\partial  \dq} = \gamma^3 M_0 >0,
\eq
thus the {\it necessary} condition for stability is fulfilled.  In this section and what follows we will confine ourselves to the case where $\kappa=1$ and also $Q=1$.  For that case from the Bogolubsky stability requirement \cite{ref:bogol} we know that without forcing when 
$\omega < \omega_c = 0.697586$, the solitary waves are unstable.

\begin{figure}[ht!]
\begin{center}
\begin{tabular}{c}
\includegraphics[width=8.0cm]{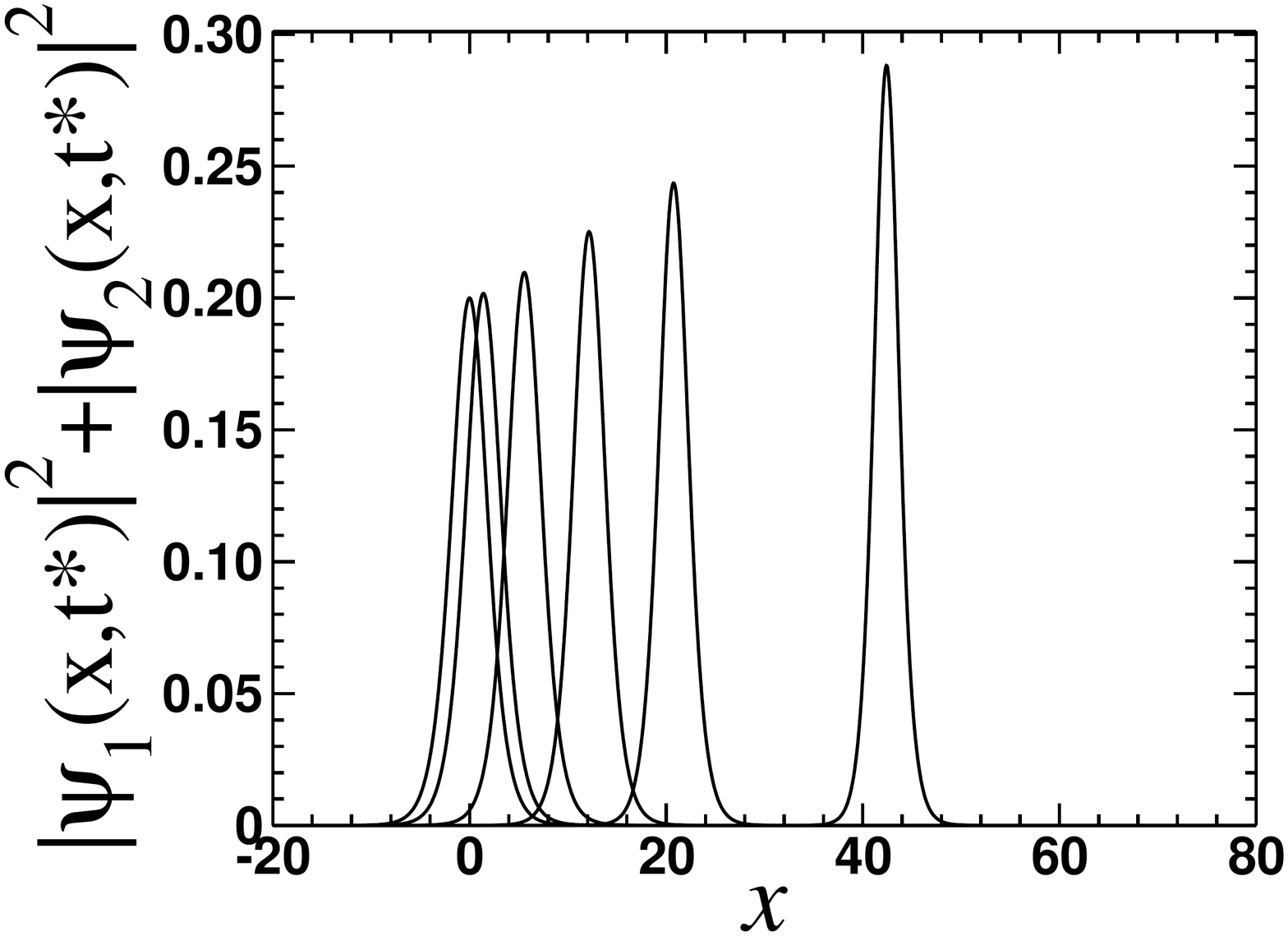}  \\
\\
\\ 
\includegraphics[width=8.0cm]{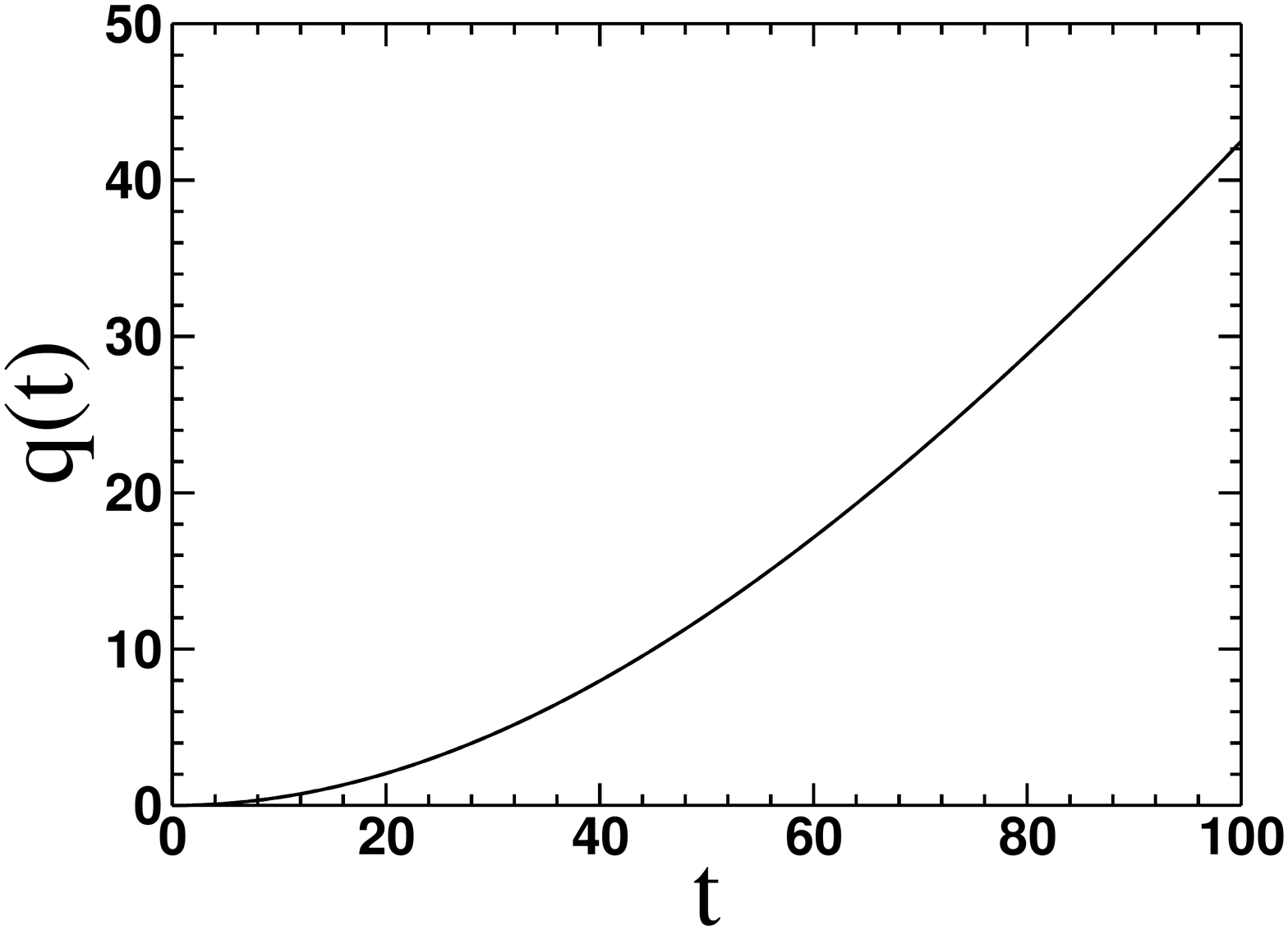} \\
\\
\\ 
\includegraphics[width=8.0cm]{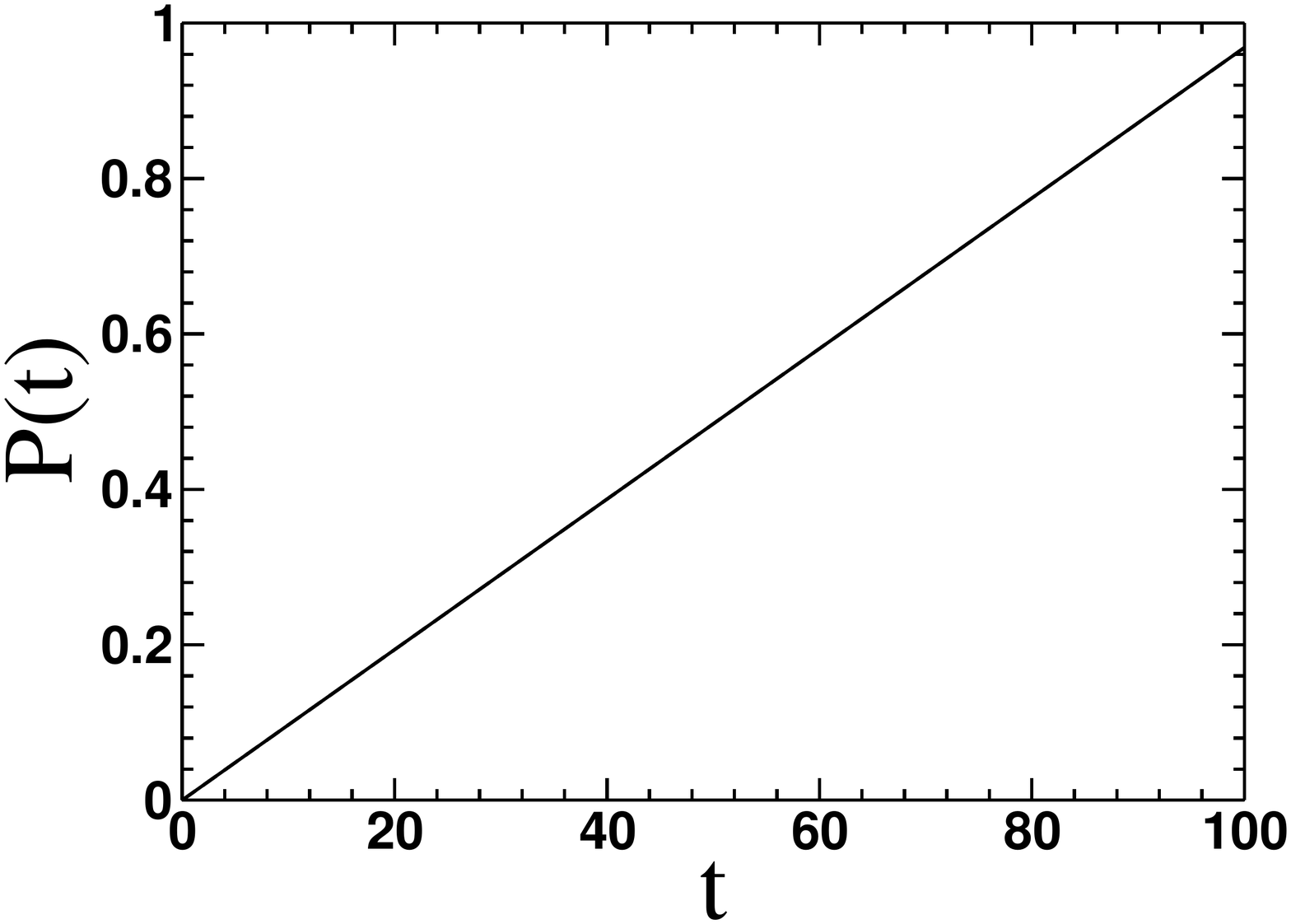} 
\end{tabular}
\end{center}
\caption{Simulations with a ramp potential, $V(x)=-V_{1} x$. 
Upper panel: Charge density $\rho_Q$ at $t^{*}=16.6; 33.3; 50; 66.6; 83.3; 100$. 
Middle and lower panels: solitary wave position $q(t)$ and momentum $P(t)$ from analytical results of the CC equations (solid lines) 
and from numerical simulations (dashed lines) of the forced NLDE. 
The curves are super-imposed. For the final time of integration the relative error of $q(t)$ is of order $10^{-5}$.   
Parameters: $g=1$, $m=1$, $\omega=0.9$ and $V_{1}=0.01$. Initial condition: exact solitary wave of the unperturbed NLDE with 
 zero initial velocity.}
\label{ramp1} 
\end{figure}

\begin{figure}[ht!]
\begin{center}
\begin{tabular}{cc}
\ & \\
\includegraphics[width=8.0cm]{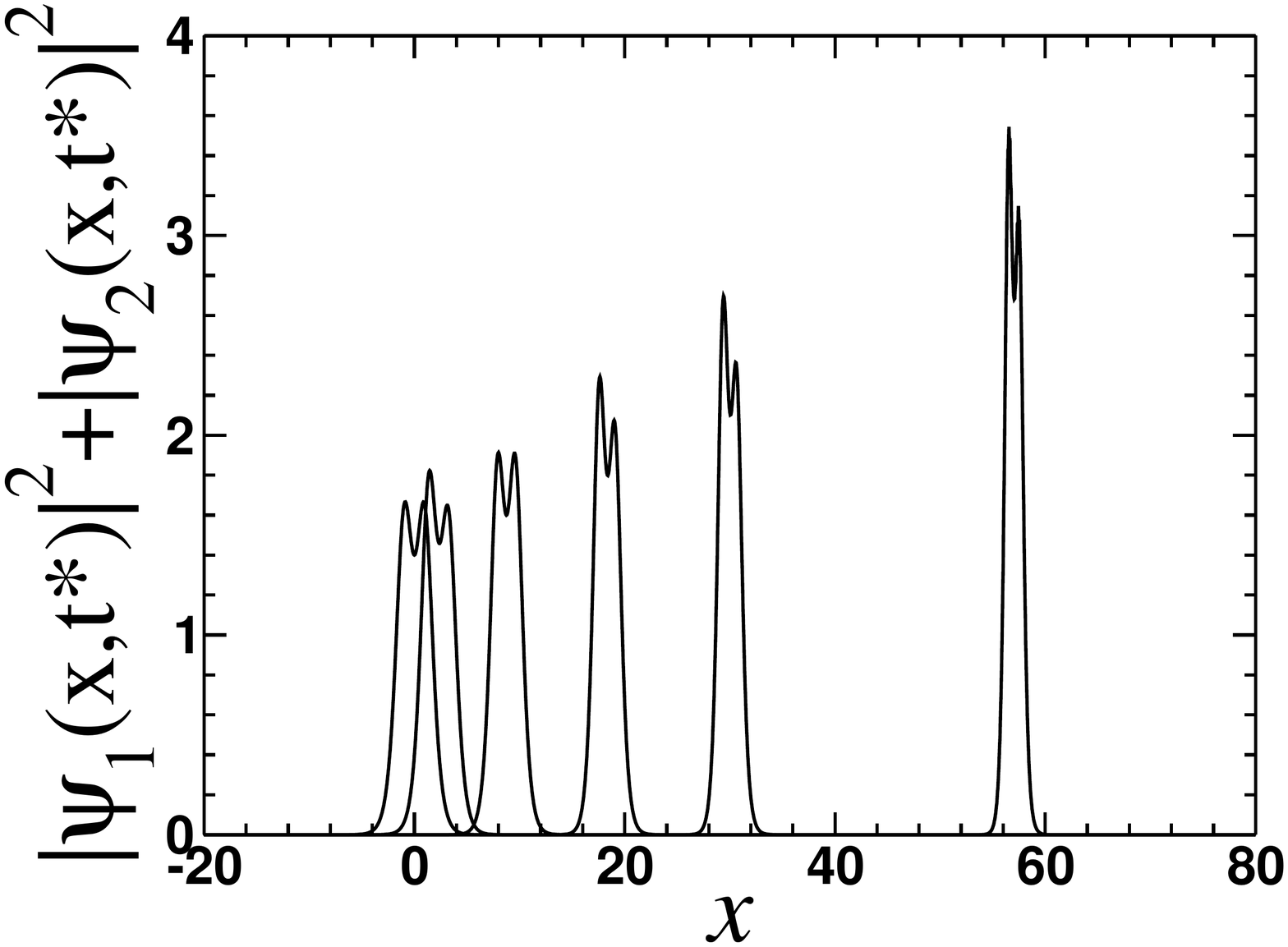}  & 
\quad \includegraphics[width=8.0cm]{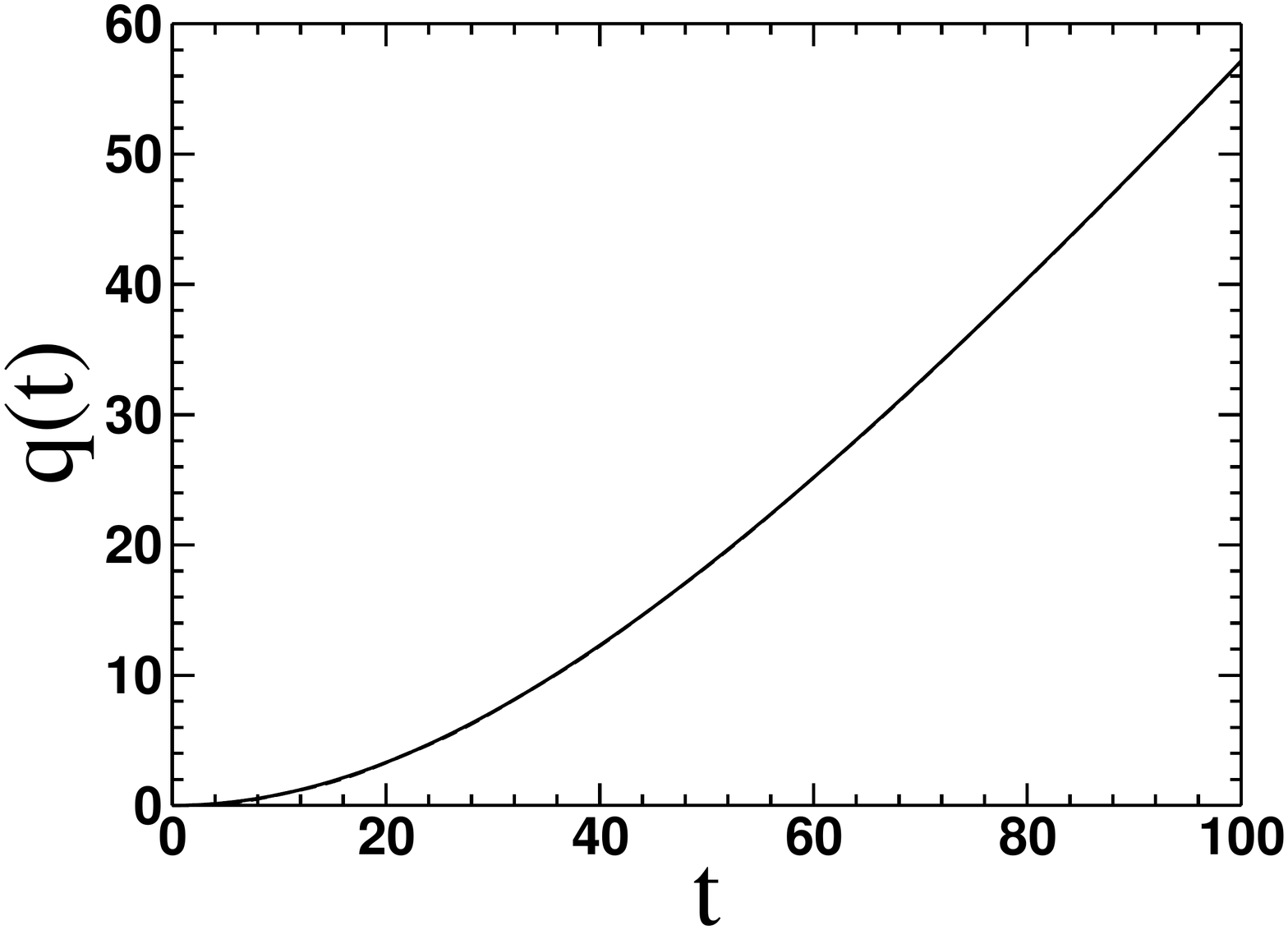} \\
 \\
\\
\\ 
\includegraphics[width=8.0cm]{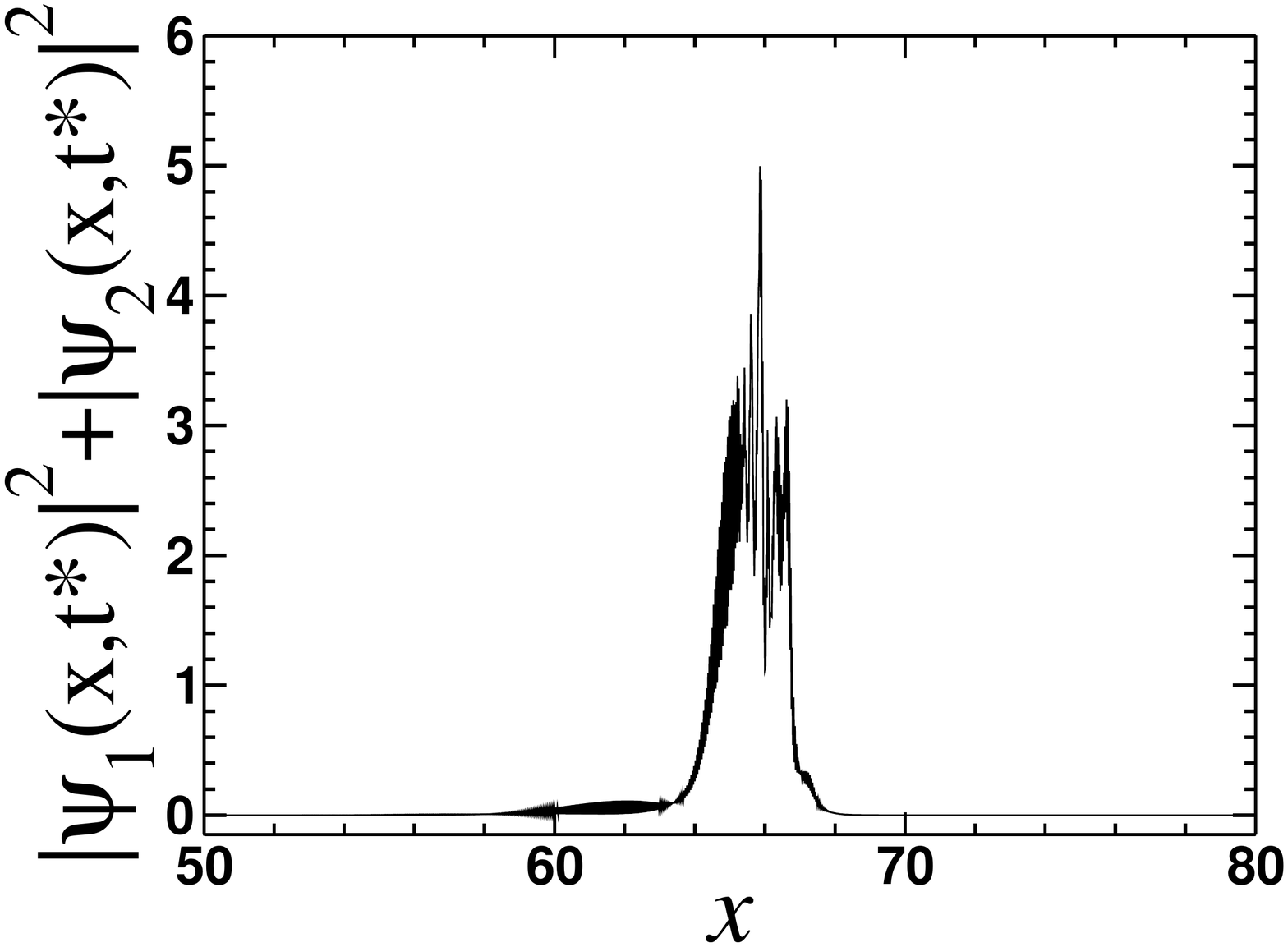}  & 
\quad \includegraphics[width=8.0cm]{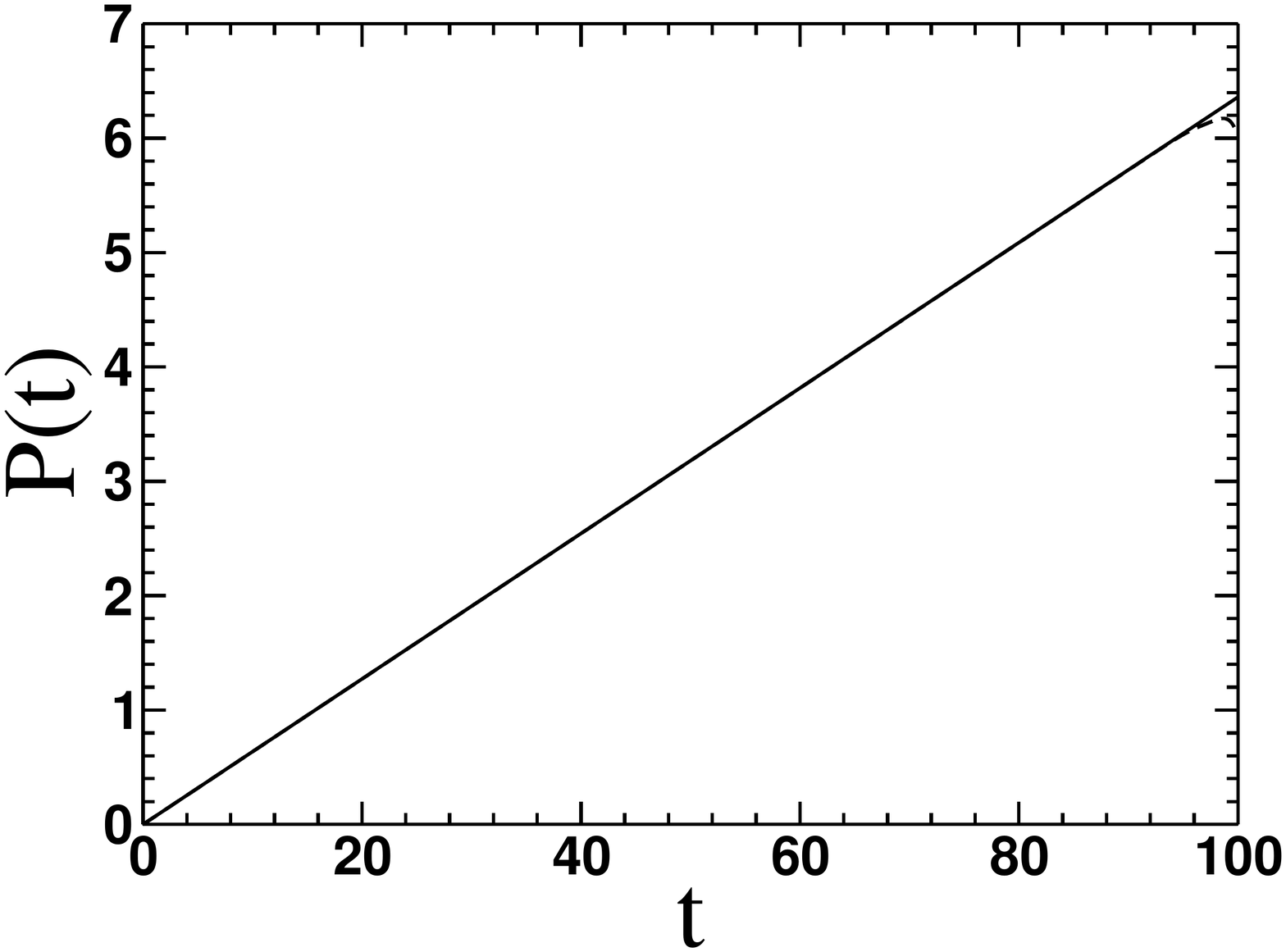}
\end{tabular}
\end{center}
\caption{Simulations with a ramp potential, $V(x)=-V_{1} x$ with $\omega$ in the unstable regime.   
Left upper panel: Charge density $\rho_Q$ at $t^{*}=16.6; 33.3; 50; 66.6; 83.3; 100$. 
Left lower panel: Charge density at $t^{*}=110$. 
Right panels: $q(t)$ and $P(t)$ from analytical results of the CC equations (solid lines) and from numerical simulations of the forced 
NLDE (dashed lines), the curves are super-imposed.  
Parameters: $g=1$, $m=1$, $\omega=0.3$ and $V_{1}=0.01$. Initial condition: exact solitary wave of the unperturbed NLDE with 
 zero initial velocity. 
}
\label{ramp2} 
\end{figure}

\begin{figure}[ht!]
\begin{center}
\begin{tabular}{c}
\includegraphics[width=8.0cm]{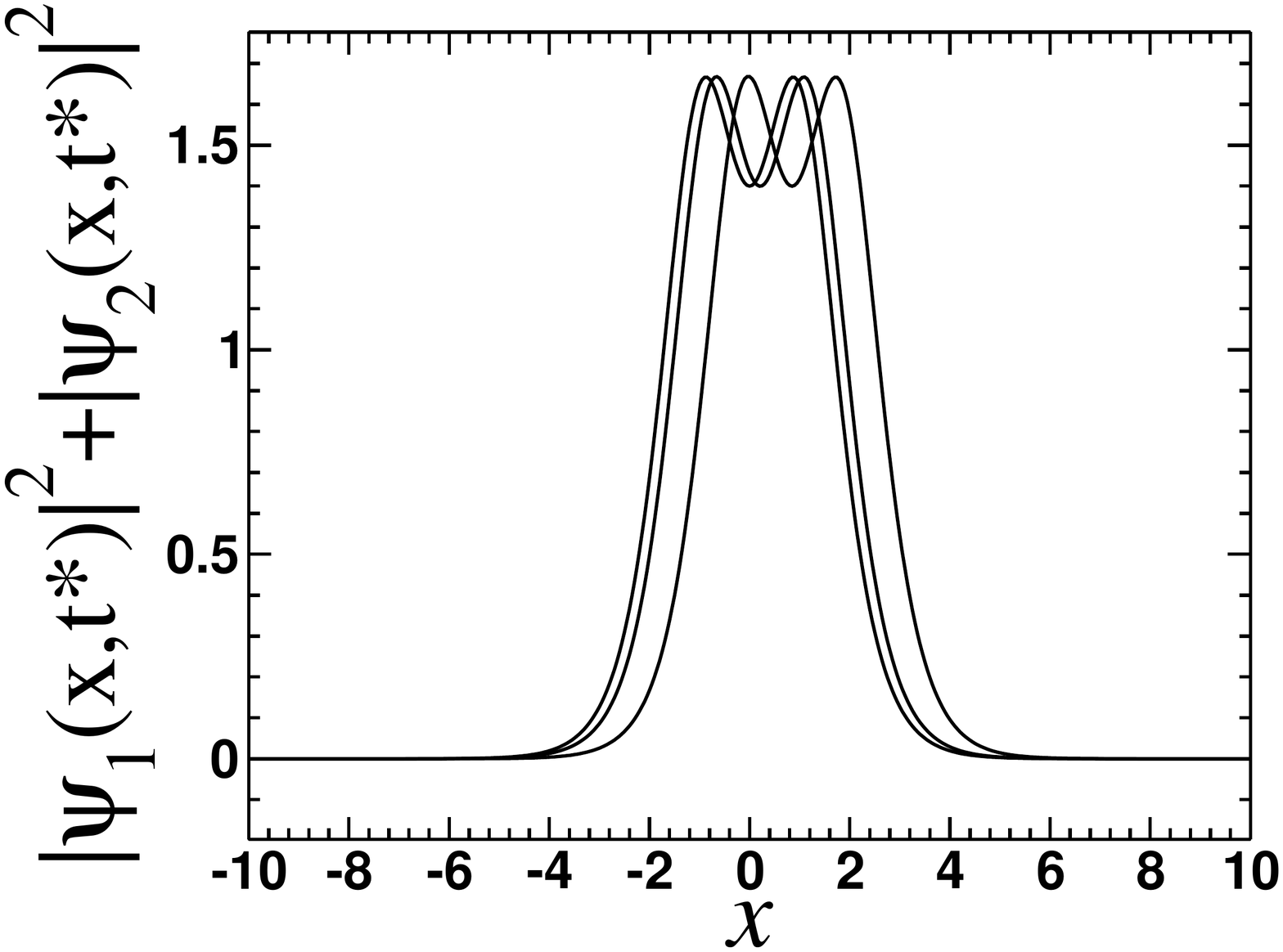}  \\
\quad \includegraphics[width=8.0cm]{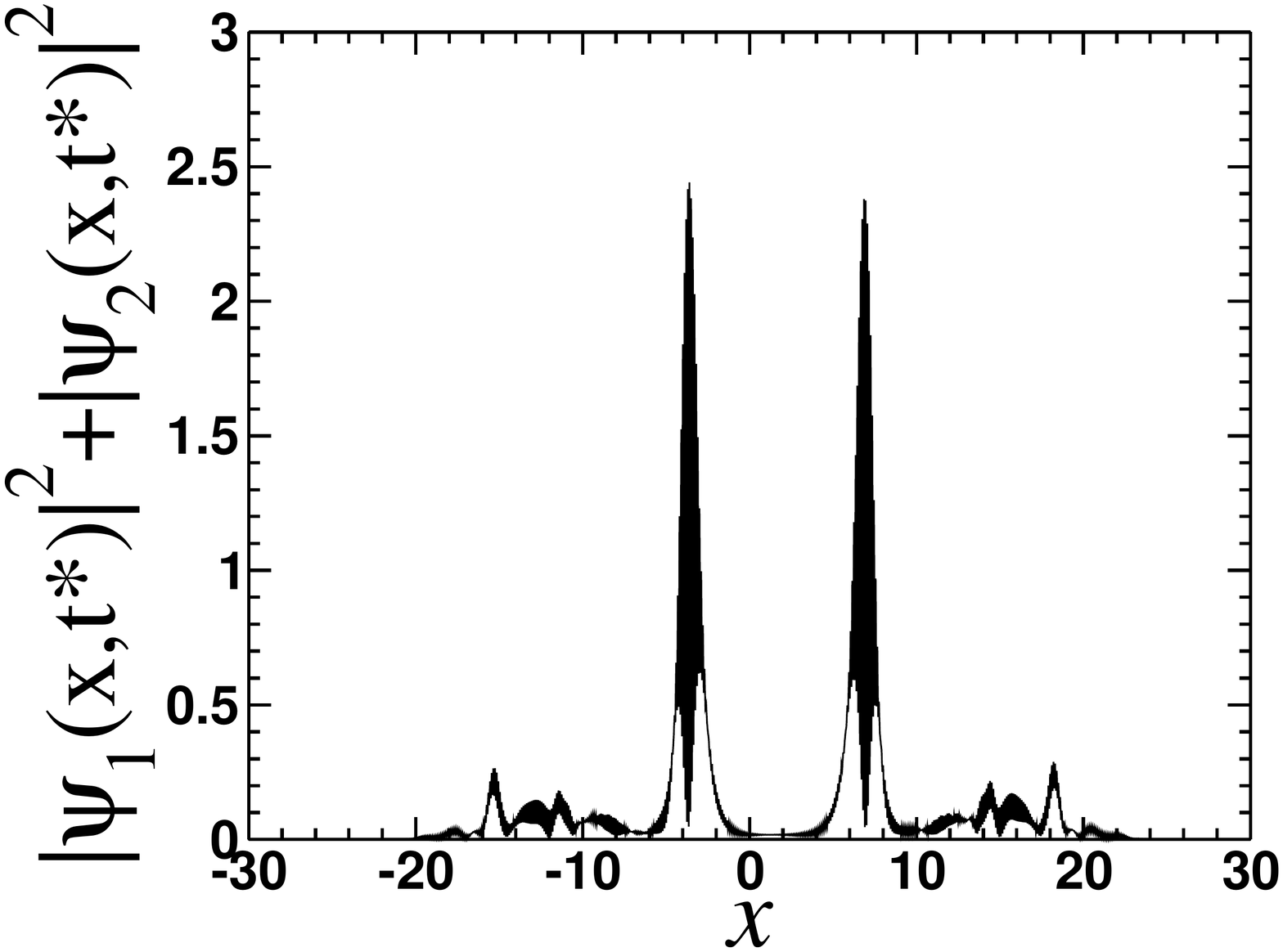} 
\\
\\ 
\includegraphics[width=8.0cm]{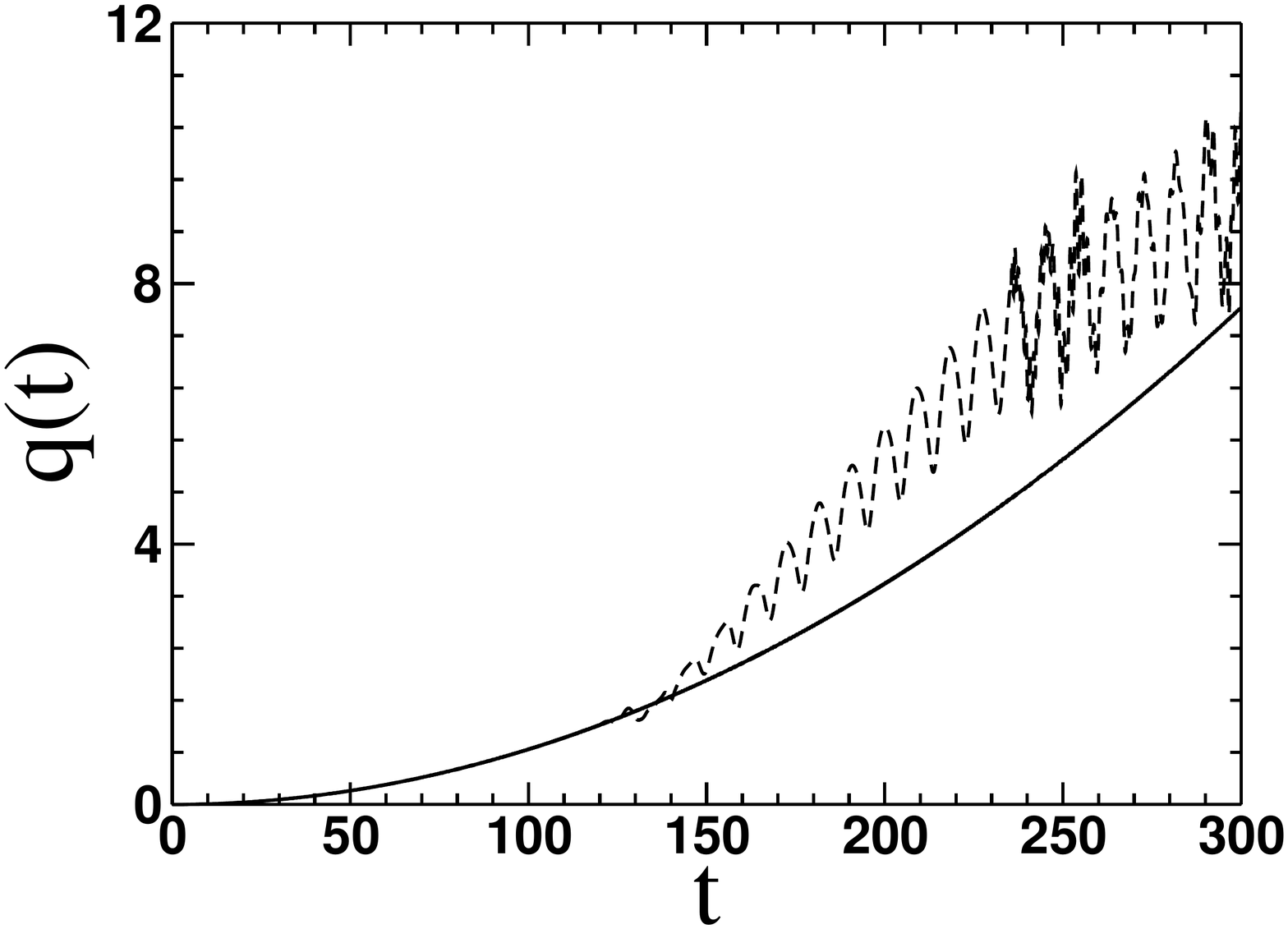}  
\end{tabular}
\end{center}
\caption{Simulations with a ramp potential, $V(x)=-V_{1} x$ with  $\omega$ in the unstable regime.  
Upper panel: Charge density $\rho_Q$ at $t^{*}=0;50;100$. 
Middle panel: Charge density at $t^{*}=150$. 
Lower panel: $q(t)$ from analytical results of the CC equations (solid line) 
and from numerical simulations (dashed line) of the forced 
NLDE.  
Parameters: $g=1$, $m=1$, $\omega=0.3$ and $V_{1}=0.0001$.  
Initial condition: exact solitary wave of the unperturbed NLDE with 
 zero initial velocity.}
\label{ramp3} 
\end{figure}

Let us look at the case where the unforced solitary wave is stable.
For $\omega=0.9, g=1$  we  have solved numerically the NLDE for $V_1=0.01$  (as well as $V_1=10^{-3}$ and $V_1= 10^{-4}$). 
We find for all these values of $V_1$  the solitary wave is stable at all simulation times and the center of the solitary wave follows the analytic formula we derived from the CC equation for $q(t)$, namely Eq. (\ref{qt}). In Fig. \ref{ramp1} we display the results of the simulation for the charge density $\rho_Q(x,t_{fixed})$, and $q(t)$, $P(t)$ for $V_1=0.01$. We notice that the width of the solitary wave gets Lorentz contracted as the velocity increases (this effect is not apparent for the smaller values of $V_1$).  Because the charge is conserved, the height of the solitary wave increases due to the increase of  $\gamma(t)$. 

For the case $\omega = 0.3$, the unforced solitary wave has double humped behavior and is unstable at late times.  Here our simulations show that until the instability sets in (around $t \approx 110$, for $m=1,g=1, V_1=0.01$) the position of the solitary wave follows the analytic solution of the CC equation Eq. (\ref{qt}).  However, the actual shape of the solitary wave becomes asymmetric with the left hump becoming higher than the right hump as a precursor to the wave becoming unstable.  This is shown in Fig. \ref{ramp2}, where $\rho_Q(x,t_{fixed})$ is plotted against $x$ for various $t=t^\star$.    In Fig. \ref{ramp3} we give results of the simulation for  the case where $\omega=0.3, V_1=0.0001$.   Here, looking at $q(t)$  we explicitly see that around $t=120$, the solution of the NLDE diverges from the solution of the CC equation.  Also for this value of the potential the solitary wave humps are symmetric and that the single solitary wave breaks up into two solitary waves with some radiation when it goes unstable. 
\subsection{Harmonic Potential}
Let us  consider the case of an external harmonic potential, 
$V(x) = \frac{1}{2} V_2 x^2$, and $V_2 > 0$.
For that case  from Eq. (\ref{udeff}) we find  that
\ba
U &&=  \frac{1}{2} V_2   \int_{-\infty}^{\infty} dz  \left[\left(q+\frac{z}{\gamma}\right)^2 [ A^2(z)+B^2(z)\right] \nonumber \\
&& =  \frac{1}{2} V_2 \left[ Q q^2(t) + \left(1- \dq^2(t) \right) I_3 \right], 
\ea
where  $ I_3 = \int_{-\infty}^{\infty} dz ~~ z^2 [ A^2(z)+B^2(z)]  $.
From Eq. (\ref{Feff}) we have
\bq
 F_{eff} [q,{\dot q} ] = - V_2 I_3 \ddot{q} - V_2 Q q , 
 \eq
 leading to the equation of motion [see Eq.(\ref{force1})]:
 \bq
 \frac{d}{dt} \left\{ \left(M_0   \gamma  + V_2 I_3 \right) \dq \right\} = - V_2  q(t) Q.
 \eq
 This can be rewritten as 
 \bq
(M_0 \gamma^3 + V_2 I_3) {\ddot q}  + V_2 Q q = 0.   \label{qddot1}
 \eq
 In the non-relativistic regime where $ \gamma \approx 1$ we recover the oscillator equation for the collective coordinate $q(t)$ namely
 \bq\label{minc}
 \ddot{q} + \Omega^2 q =0, ~~~ \Omega^2 = \frac{V_2 Q}{M_0+V_2 I_3}.
 \eq
Note that the rest mass is {\it increased} by the term $V_2 I_3 > 0$.
For initial conditions $q(0)=0$, $\dq(0)= v_0$  we obtain
\bq
q(t) = \frac{v_0}{\Omega}  \sin \Omega t . 
\eq
\subsubsection{Energy conservation}
From the energy conservation equation (\ref{econs}) 
we obtain
\bq
E =  M_0 \gamma  + \frac{1}{2} Q V_2 q^2 +  \frac{1}{2} V_2 I_3 (1 + \dq^2) .
\eq

In the low velocity limit,  we need to keep the first two terms in the expansion of $\gamma$ in the expression for the energy;
namely  (here we suppress the speed of light in the $v/c$ expansion)
\bq
\gamma = 1 + \frac{1}{2} \dq^2 + ... ~. 
\eq

We then have for the solution $
q(t) = \frac{v_0}{\Omega}  \sin \Omega t $ the usual equipartition of energy and that the non-relativistic energy  is twice the initial kinetic energy (apart from a constant)
\bq
E =  M_0+ 2 T,  ~~~ T= (M_0+V_2 I_3) v_0^2
\eq
with the effective mass of the solitary wave increased over the unforced case by the quantity $V_2 I_3$.
\subsubsection{Canonical momentum and stability criterion}
From the equation for the canonical momentum, Eq. (\ref{canp}), we find
\bq
P= (M+ V_2 I_3) \dq = (M_0 \gamma + V_2 I_3) \dq . \label{canp2}
\eq
This again shows the mass increased by $V_2 I_3$. The stability criterion Eq. (\ref{pdot}), leads to 
\bq \label{pdotharmonic} 
\frac{\partial P(q,\dq)}{\partial {\dot q}}  =\gamma^3 M_0 + V_2 I_3 >0.
\eq 
Thus the necessary condition for stability of the solitary wave is fulfilled. 
 
We would now like to see how well the CC equations for $q$ and $P$, namely Eqs. (\ref{qddot1}) and (\ref{canp2}), compare with the numerical solutions of the forced nonlinear Dirac equation.   We will choose our initial condition to be $q(0)=0, \dq(0) = v_0$ and study both
 the non-relativistic regime ($v_0=0.1$) and the relativistic regime ($v_0=0.9$).   The external potential can be written as
 \bq
 V = \frac{1}{2} V_2 x^2 =  \frac{1}{2}\left(\frac{x}{l}\right)^2 , 
 \eq
 which identifies the characteristic length of the potential as $l = 1/\sqrt{V_2}$.  We would like to choose the characteristic length of the potential to be large compared to the width of the solitary wave which is $1/(2 \beta)$, $\beta = \sqrt{1-\omega^2}$.  Choosing
 $V_2 = 10^{-4}$ accomplishes this requirement.  At low velocities both $P(t)$ and $\dq(t)$ are proportional to $\cos \Omega t$, thus 
 $P(\dq)$ is a straight line with positive slope. 
 
 First let us consider the regime where the unforced solitary waves are stable, and choose $\omega=0.9, g=1$.  In the non-relativistic regime ($v_0=0.1$) we get the results shown in Fig. \ref{harm1}. The oscillations of both $q(t)$ and $P(t)$ are harmonic as predicted by 
 the CC equations. The charge density maintains its shape as its position oscillates periodically in time. 
For $v_0=0.9$,  $q$ and $P$  again follow the CC equations for a little less than half the oscillation period but then the solitary wave becomes unstable  and the exact simulation of  $q$ and $P$ then diverges from the solution to the CC equations. This is shown in Fig. \ref{harm2}. 

Next we consider the regime $ 0 < \omega < \omega_c=0.697586$  where the unforced solitary wave is unstable. 
For the parameters   $\omega=0.3, g=1, V_2 = 10^{-4}$, we obtain the typical result found in the unforced problem that  at around $t=120$, the solitary wave becomes unstable. Until then the CC equations for $q(t)$ and $P(t)$ track well the exact solution. This is seen in Fig. 
\ref{harm3}. However, the wave function starts becoming asymmetric at late times and departs from our symmetric ansatz even before the solitary wave becomes unstable and breaks into two solitary waves plus some radiation. 
 
\begin{figure}[ht!]
\begin{center}
\begin{tabular}{c}
\includegraphics[width=8.0cm]{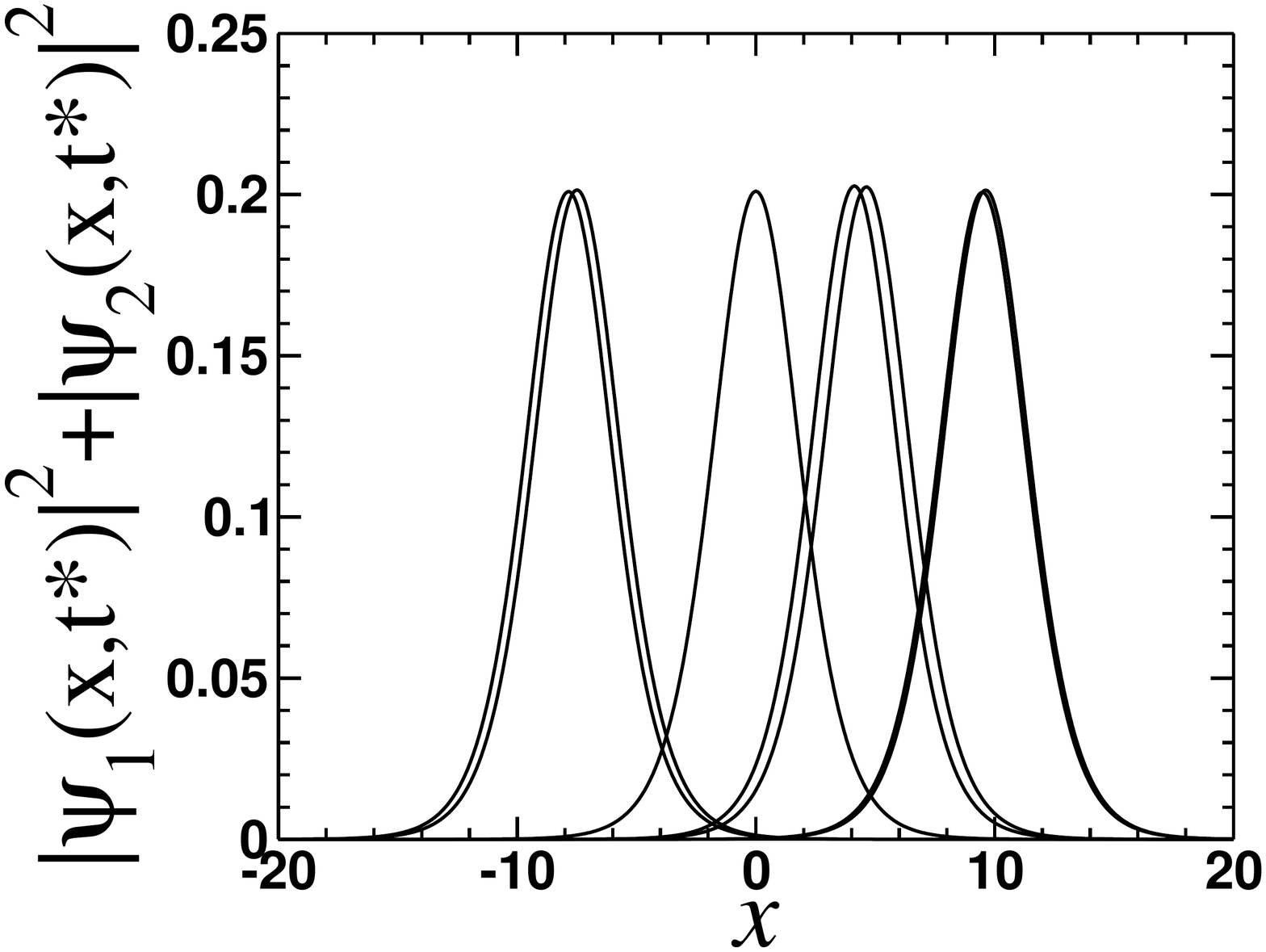}  
\\
\\
\\
\includegraphics[width=8.0cm]{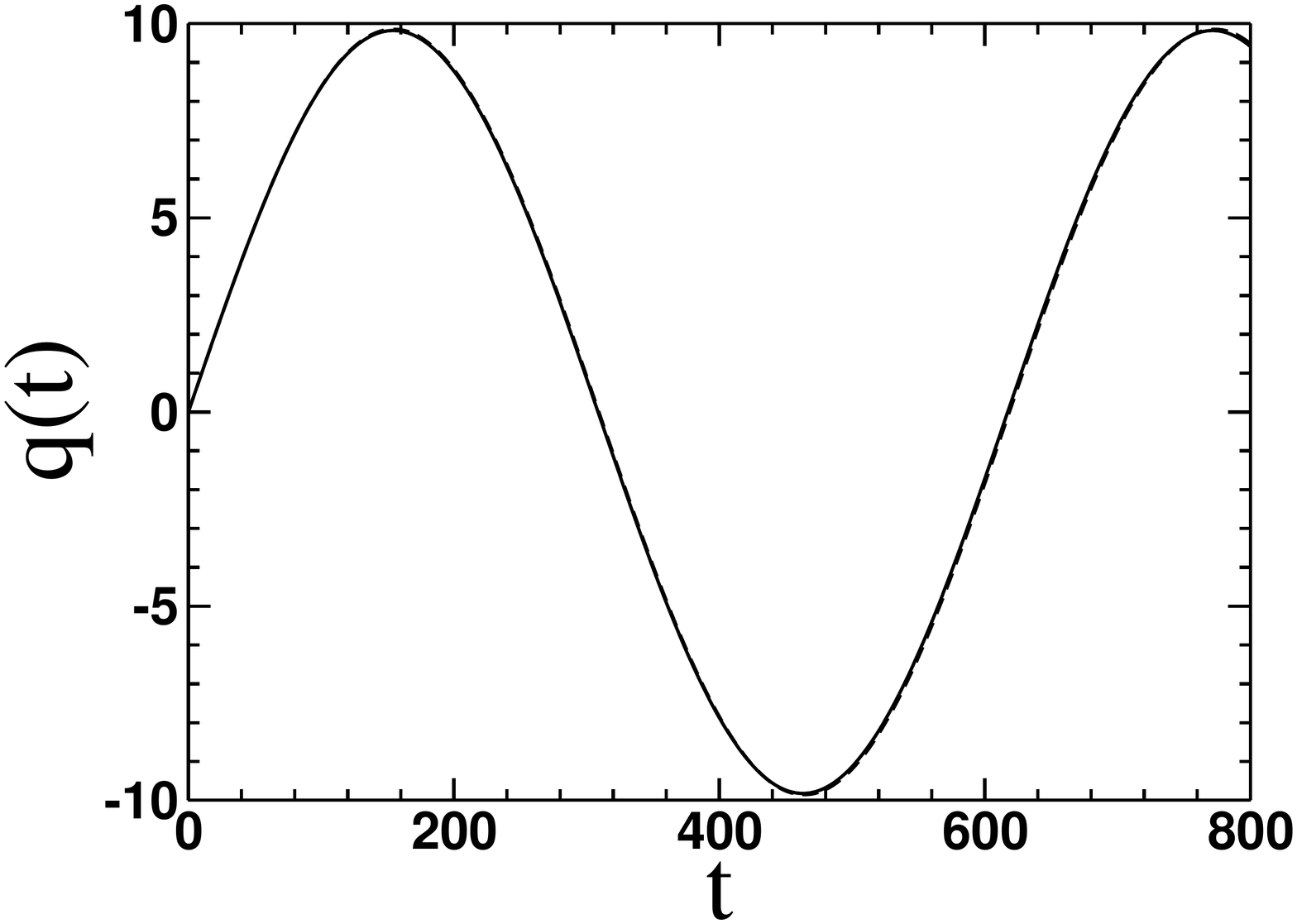}  \\
\\
\\
\includegraphics[width=8.0cm]{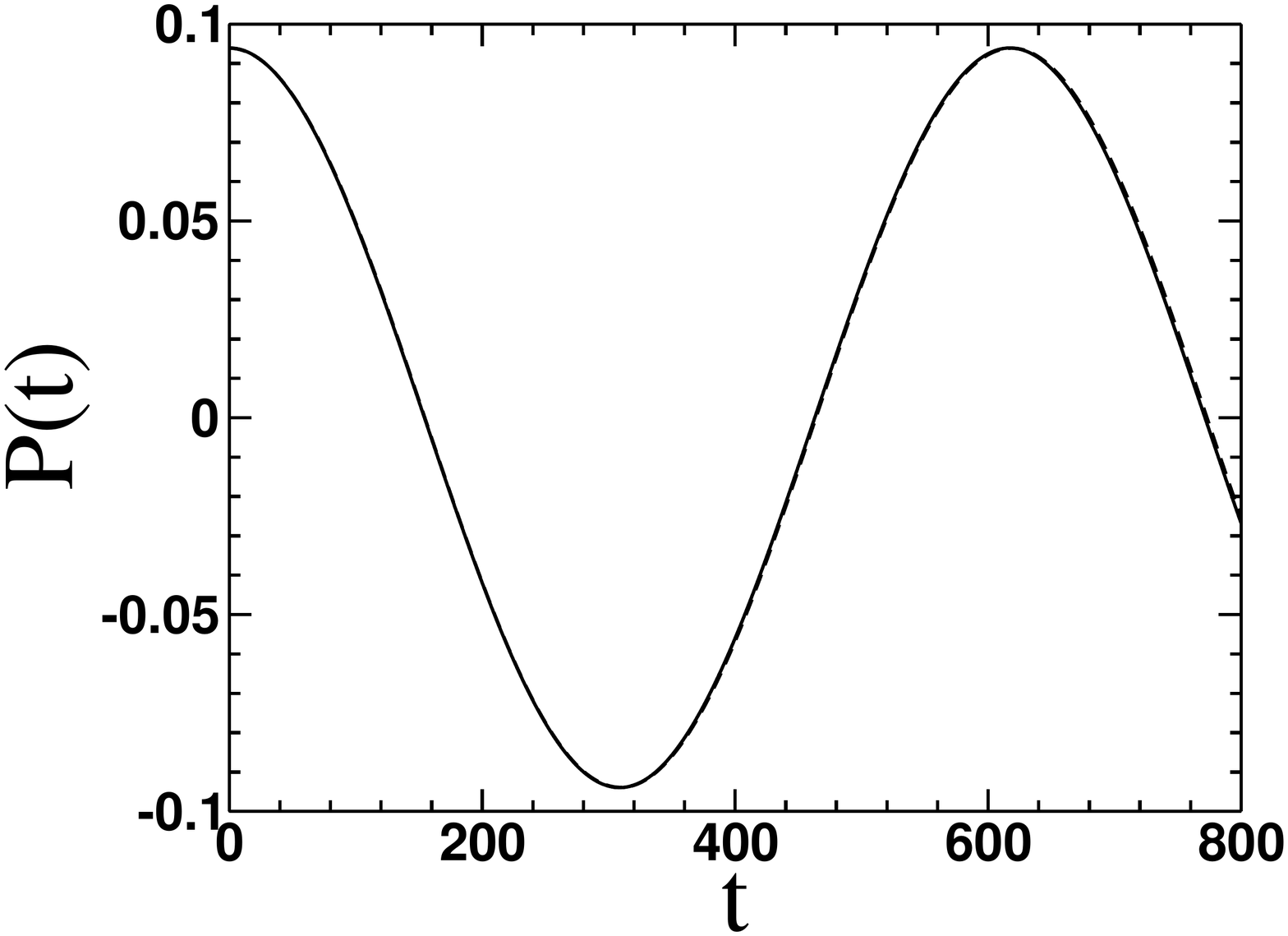}  
\end{tabular}
\end{center}
\caption{Harmonic potential, $V(x)=(V_{2}/2) x^{2}$. Upper panel: Charge density $\rho_Q$ at 
$t^{*}=0;133.3;266.6;400;533.3;666.6;800$. 
Middle and lower panels: solitary wave position $q(t)$ and momentum $P(t)$, from the numerical solutions of the CC equations (solid lines) 
and from numerical simulations (dashed lines) of the forced 
NLDE.  The curves are super-imposed. 
The charge 
$Q=0.96864$ and energy $E=0.93921$ are both conserved. 
Parameters: $g=1$, $m=1$, $\omega=0.9$ and $V_{2}=10^{-4}$. 
Initial condition: exact solitary wave of the unperturbed NLDE with 
 initial velocity $v(0)=0.1$.}
\label{harm1} 
\end{figure}


\begin{figure}[ht!]
\begin{center}
\begin{tabular}{c}
\includegraphics[width=8.0cm]{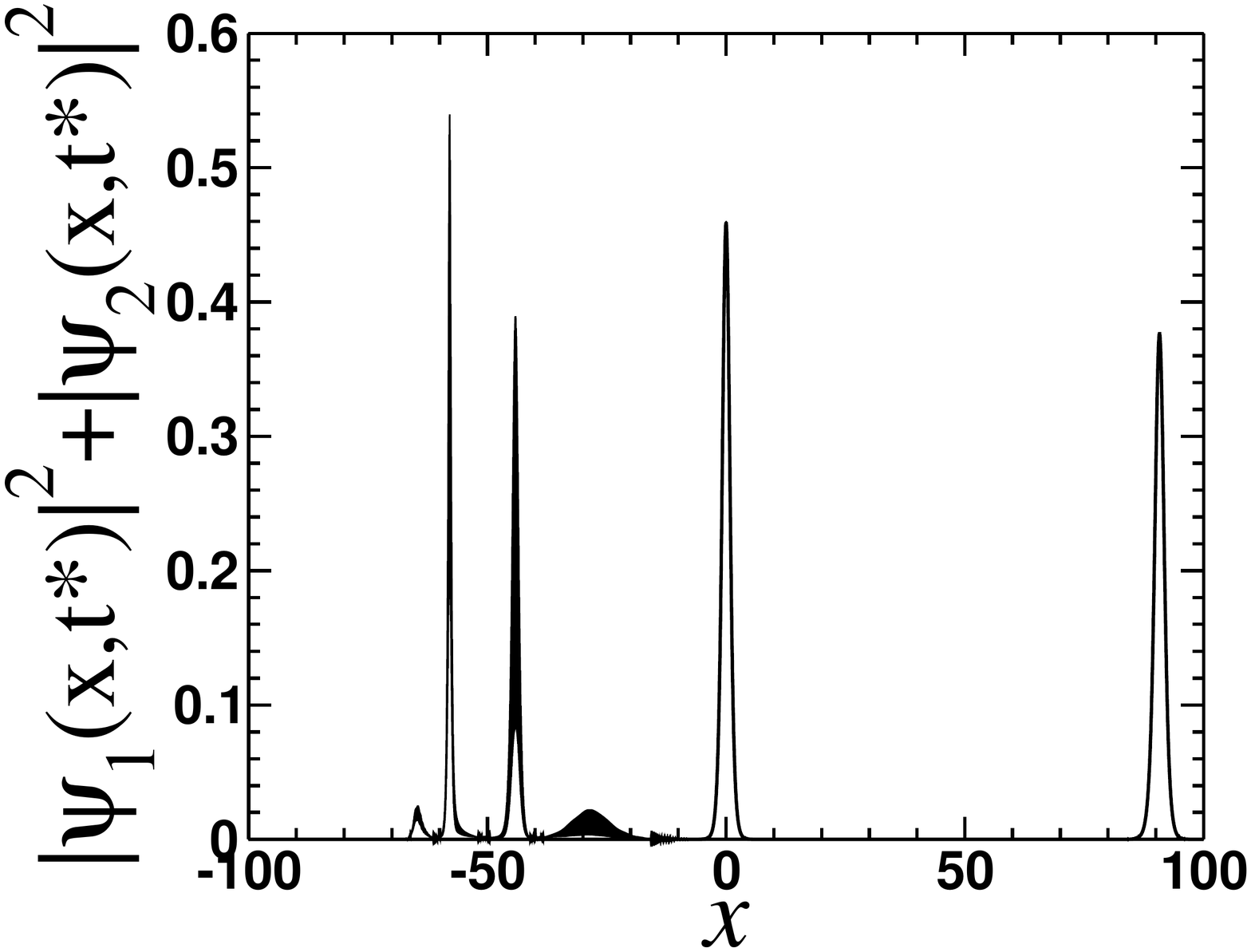}
\\
\\
\\
\includegraphics[width=8.0cm]{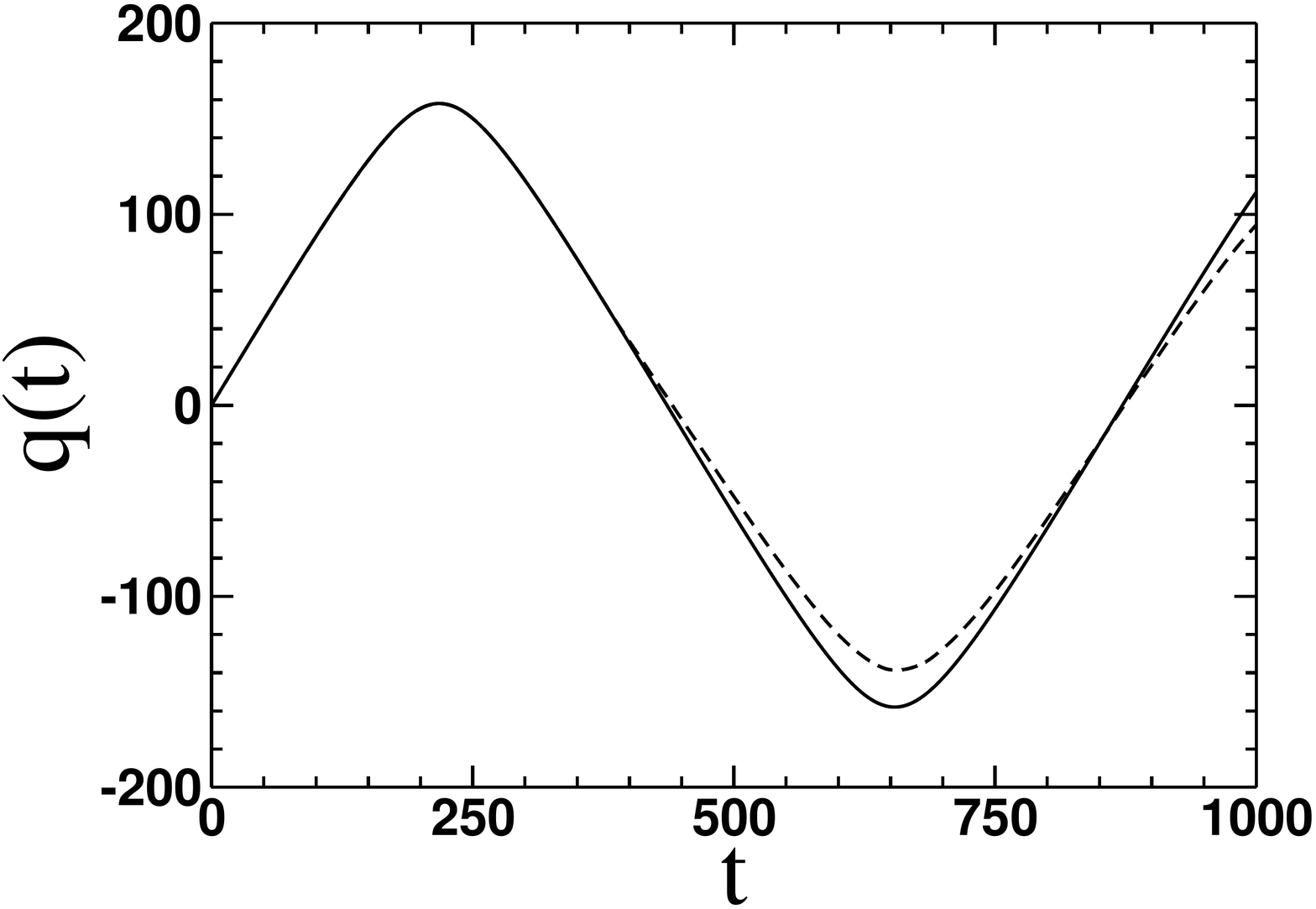} \\
\\
\\
\includegraphics[width=8.0cm]{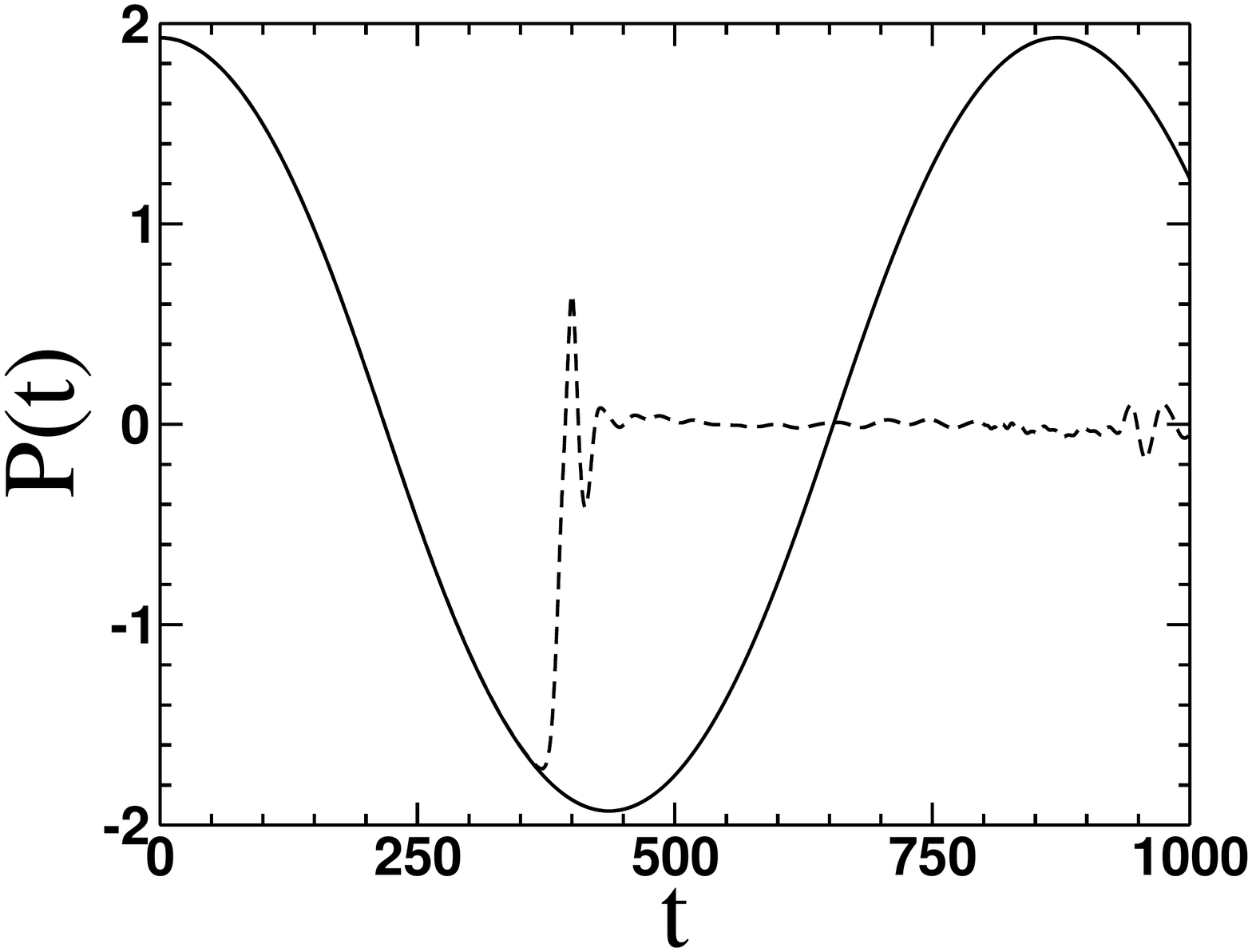}  
\end{tabular}
\end{center}
\caption{Harmonic potential, $V(x)=(V_{2}/2) x^{2}$. Relativistic regime, unstable solitary wave. Upper panel: Charge density $\rho_Q$ at 
$t^{*}=0;133.3;500$. 
Middle and lower panels: solitary wave position $q(t)$ and momentum $P(t)$, from the numerical solutions of the CC equations (solid lines) 
and from numerical simulations (dashed lines) of the forced 
NLDE.  
 The curves in $P(t)$ are super-imposed only till $t=370$.  
Parameters: $g=1$, $m=1$, $\omega=0.9$ and $V_{2}=10^{-4}$. 
Initial condition: exact solitary wave of the unperturbed NLDE with 
 initial velocity $v(0)=0.9$.}
\label{harm2} 
\end{figure}

\begin{figure}[ht!]
\begin{center}
\begin{tabular}{cc}
\ & \\
\includegraphics[width=8.0cm]{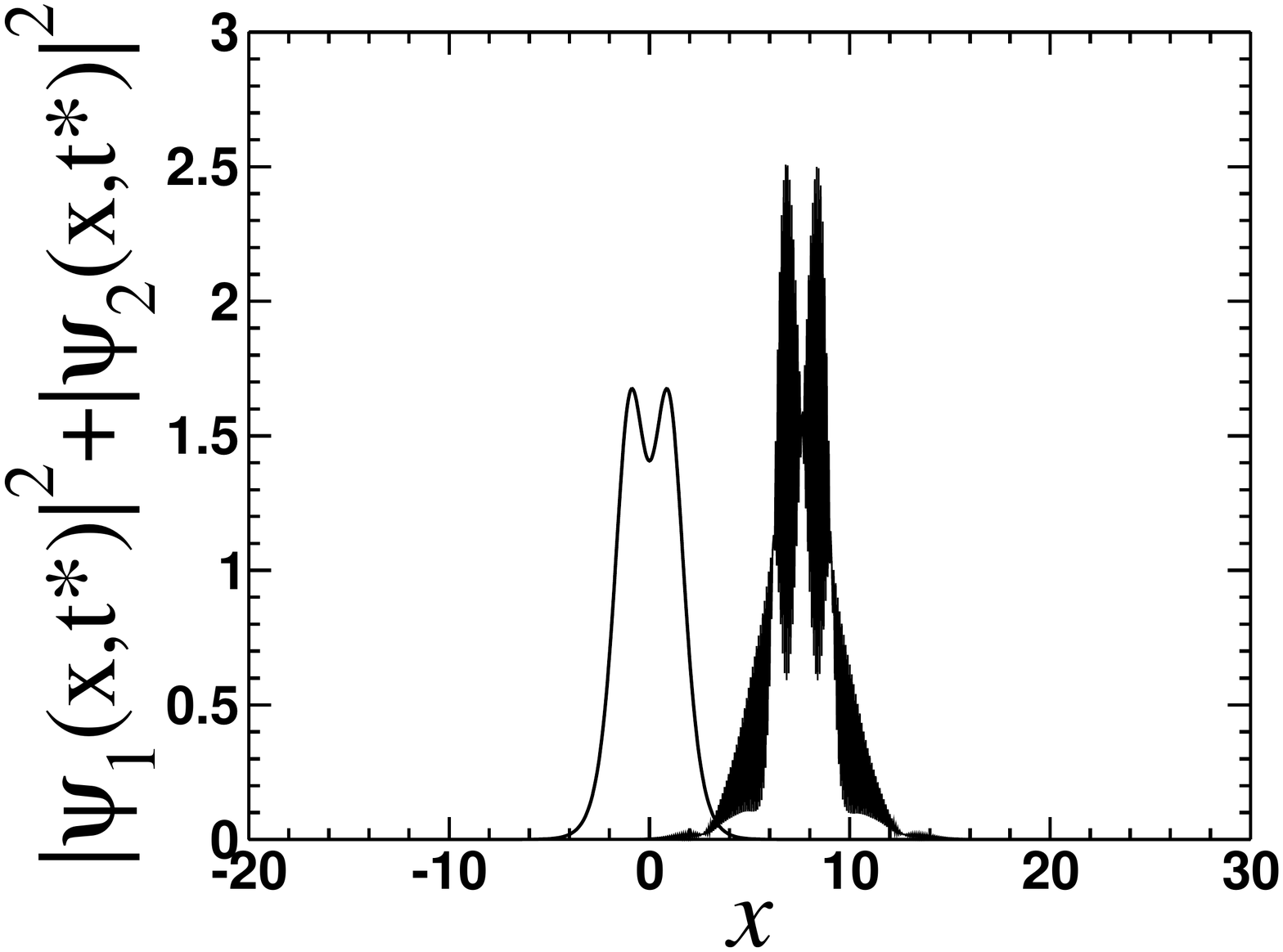}  & 
\quad \includegraphics[width=8.0cm]{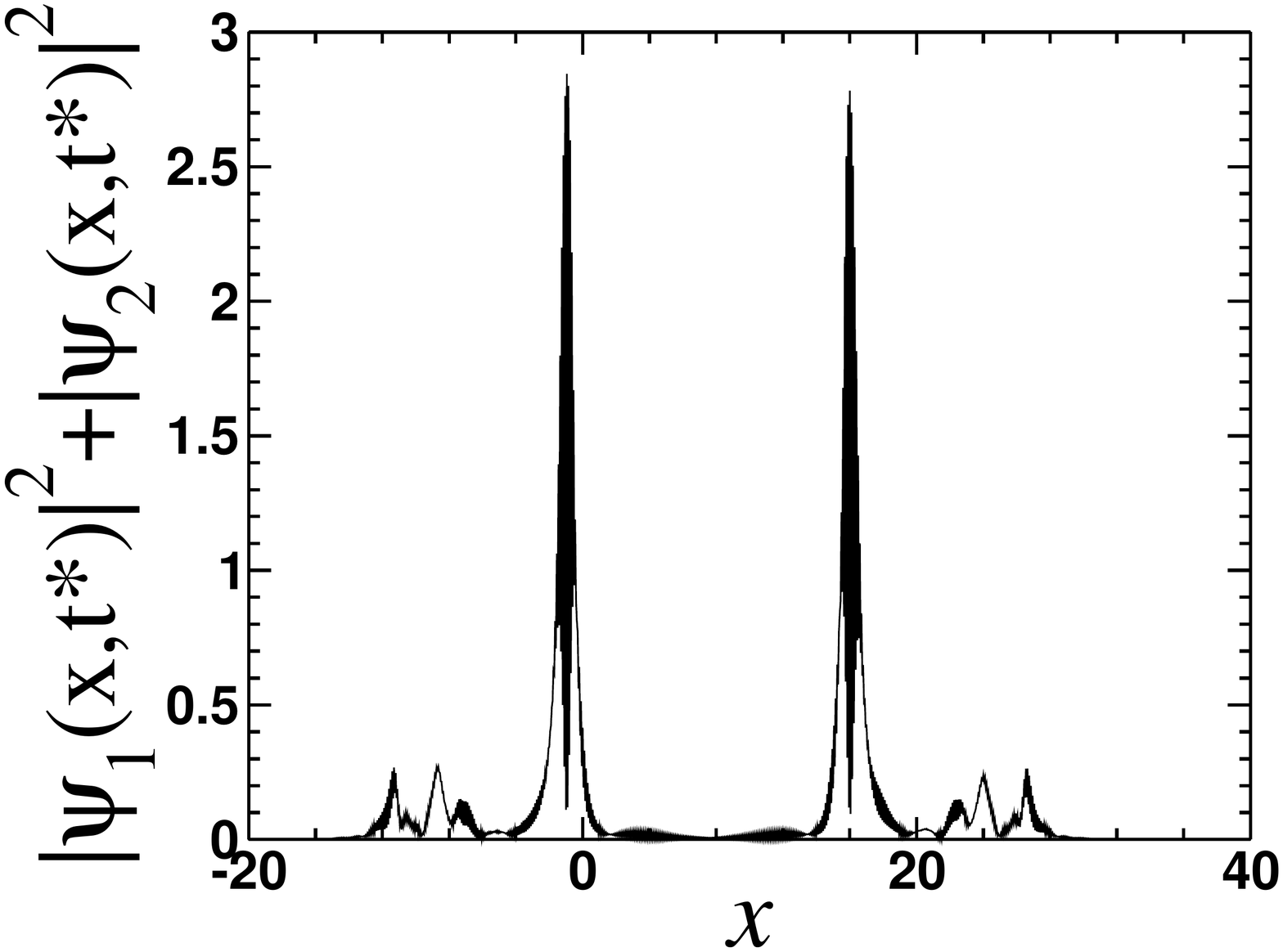} 
\\
\\
\\ 
\includegraphics[width=8.0cm]{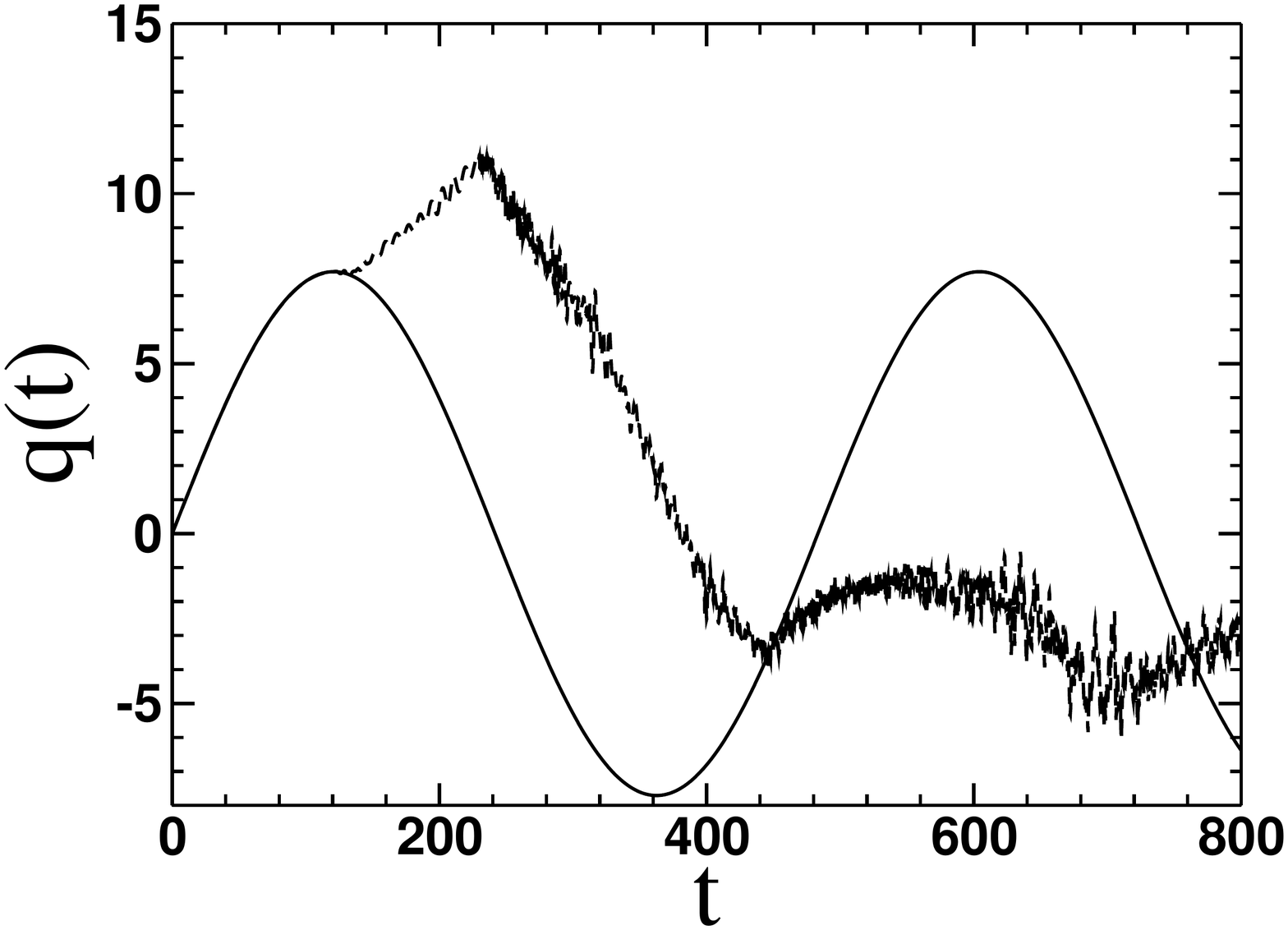}  & 
\quad \includegraphics[width=8.0cm]{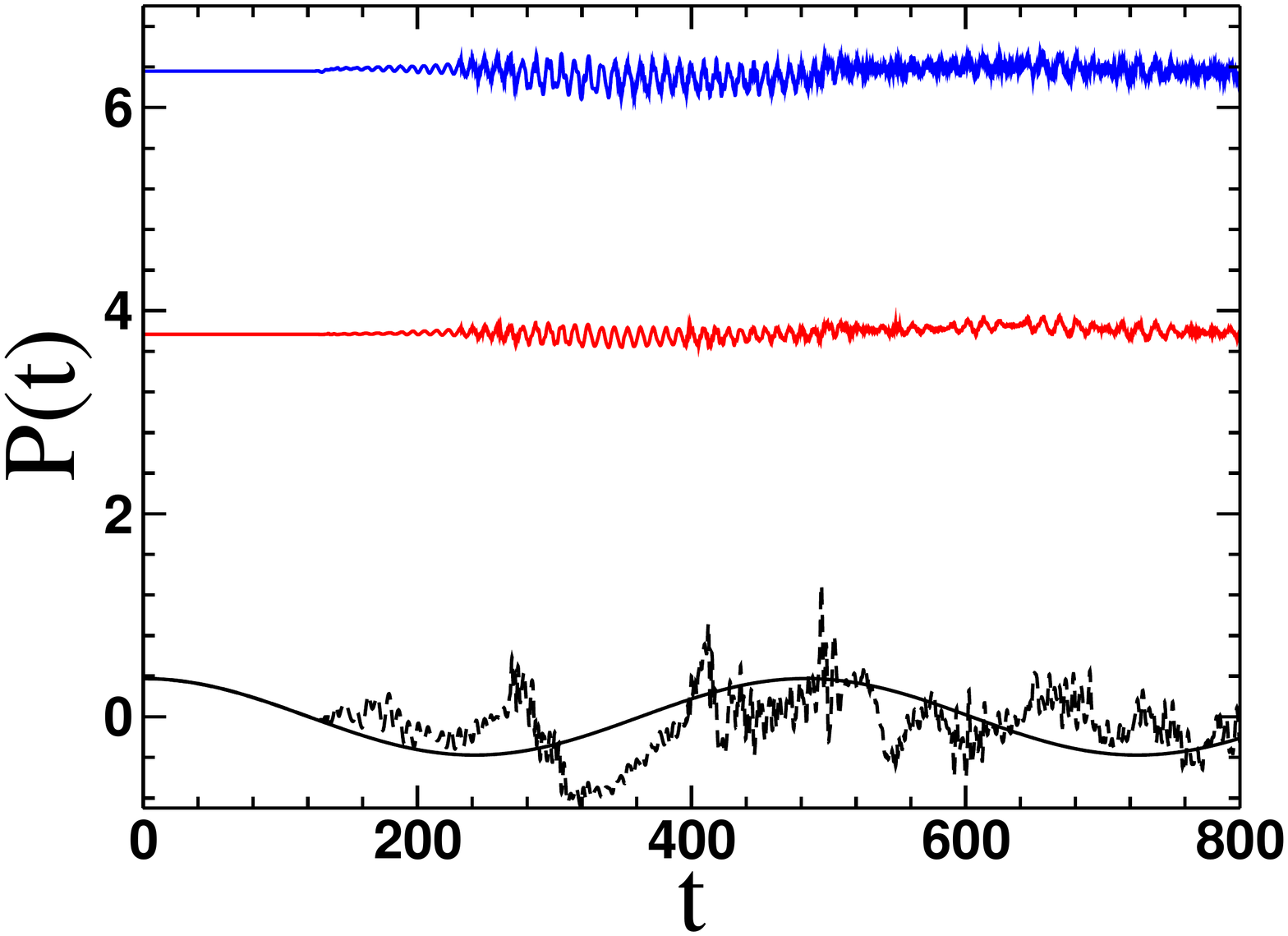}  
\end{tabular}
\end{center}
\caption{(Color online). Harmonic potential, $V(x)=(V_{2}/2) x^{2}$ with $\omega$ in the unstable regime. 
Left upper panel: Charge density $\rho_Q$ at 
$t^{*}=0;133.3$.  
Right upper panel: Charge density at 
$t^{*}=150$.  
Lower panels: solitary wave position $q(t)$ and momentum $P(t)$, 
from the numerical solutions of the CC equations (black solid lines) 
and from numerical simulations (dashed lines) of the forced 
NLDE. 
 The energy 
(red or middle curve) and charge (blue or upper curve) are also plotted.  
Parameters: $g=1$, $m=1$, $\omega=0.3$ and $V_{2}=10^{-4}$. 
Initial condition: exact solitary wave of the unperturbed NLDE with 
 initial velocity $v(0)=0.1$.}
\label{harm3} 
\end{figure}

\newpage
 \subsection{Spatially periodic potentials}
 Next consider a spatially periodic potential  
 \bq
 V(x) = - \epsilon \cos kx, ~~ \epsilon >0,
 \eq
 where the spatial period $L =  2 \pi/k  \gg 1/\beta$, and
 $1/\beta$ is the width $b$  of the solitary wave.  The potential is then a function of $q, \dq$ and is given by
 \bq
 U[q,\dq] = -\epsilon \cos k q \int dz \cos \frac{k z}{\gamma}  
\left[A^2[z]+B^2[z] \right] =  -\epsilon \cos k q\,  I_4[\dq] , 
 \eq
 so that
 \bq
 \frac{\partial U}{\partial q} =   k \epsilon \sin k q I_4[\dq] , 
 \eq
 \bq
 \frac{\partial U}{\partial \dq} =  -k  \epsilon   \gamma  \dq \cos k q 
\int dz~ z  \sin \frac{k z}{\gamma} 
\left[A^2[z]+B^2[z] \right] \equiv  -k \epsilon   \gamma  \dq \cos k q \, I_5 [\dq] . 
 \eq
 
  The generalized force Eq. (\ref{Feff})   can be written as
 \bq
F_{eff} [q,\dq] =  -k \epsilon   {I_4}\left( \dq 
\right) \sin (k  {q}(t)) - k \epsilon \cos k q ~\ddot{q} \left[ \gamma^3 I_5[\dq] -  k (\gamma \dq)^2 I_6[\dq] \right] = M_0 \frac{d}{dt} ( \gamma \dq) ,  \label{forceg}
 \eq
 where
\bq
I_6[\dq] = \int dz~ z^2 \cos 
 \frac{k z}{\gamma} 
\left[A^2[z]+B^2[z] \right] .
\eq

 In the non-relativistic limit  $(\dq)^2 \ll 1$,  $\gamma \approx 1$ and we obtain for the force law:

  \bq \label{eq733}
  (M_0 + k \epsilon I_5^0 \cos k q) \ddot{q} + k \epsilon I_4^0 \sin k q =0 , 
\eq
 where
 \bq  \label{i45}
 I _4^0= \int dz ~ \cos {k z}  \left[A^2[z]+B^2[z] \right] ;~~I_5^0= \int dz~ z  \sin {k z} 
\left[A^2[z]+B^2[z] \right] . 
 \eq
 When the potential is weak ($ \epsilon \ll1$) then $k \epsilon I_5^0 \ll M_0 $ and we obtain the pendulum  equation
 \bq
 M_0 \ddot{q} + k \epsilon I_4^0 \sin k q  =0. 
\eq
Letting $ C = k \epsilon I_4^0/M_0$, the solutions  are given by 
\bq
q(t)= \frac{2}{k}  \text{am}\left(\frac{1}{2} \sqrt{k} 
\sqrt{2 C t^2+k c_1
   t^2+4 C c_2 t+2 k c_1 c_2 t+2 C c_2^2+k c_1 c_2^2},~\frac{4 C}{2 C+k
   c_1}\right) , 
   \eq
   where $c_1,c_2$ are integration constants to be determined by the initial conditions, $q(0) =0; \dq(0) = v_0$.
   Here 
   \bq
   \text{am} [u,l] =  \text{JacobiAmplitude} [u,l] , 
   \eq
  where the modulus parameter $l$  (usually denoted by $m$) is 
   $l= \frac{4 C}{2 C+k  c_1}$. For the above initial conditions we find:
   \bq
   q(t) = \frac{2}{k} \text{am} \left[\frac{k v_0 t}{2}, \frac{4 C}{k v_0^2}\right] 
   = \frac{2}{k} \sin^{-1}\text{sn} \left[\frac{k v_0 t}{2}, \frac{4 C}{k v_0^2}\right] 
    \label{qccnr} .
   \eq
   \subsubsection{Energy conservation}
   From Eq. (\ref{econs}) we have that the solitary wave energy is given by 
   \bq
   E = \gamma M_0 - \epsilon \cos kq I_4[\dq] + \epsilon \cos k q ~ k \gamma \dq^2 I_5 [\dq].
   \eq
   In the non-relativistic limit we obtain
   \bq
   E= M_0 - \epsilon \cos kq I_4^0 + \left( \frac{M_0}{2} + \epsilon k I_5^0 \cos kq \right) \dq^2 . 
   \eq
   In the case of a weak potential (except for $M_0\rightarrow0$ when $\omega \rightarrow 1$)  
   \bq
   E = \left( 1 + \frac{\dq^2}{2}\right) M_0 - \epsilon \cos kq I_4^0 .
  \eq
  \subsubsection{Solitary wave momentum and dynamical stability}
  The solitary wave momentum is given by Eq. (\ref{canp}) and becomes
  \bq
  P = \gamma \left(M_0 + k \epsilon I_5[\dq] \cos kq \right) \dq.
  \eq
  In the non-relativistic regime we obtain
  \bq
  P =  \left(M_0 + k \epsilon I_5^0  \cos kq \right) \dq.
  \eq
 The  necessary  condition for stability
 \bq
 \frac{dP}{d \dq} =  \left(M_0 + k \epsilon I_5^0  \cos kq \right) >0
 \eq
 is satisfied except in the regime where $M_0 \rightarrow 0$ which is when $\omega \rightarrow 1$.  In that regime the solitary wave is very broad and the condition  $2 \pi /k \gg b $ is not fulfilled.
 
 \subsubsection{Numerical results for $q(t)$ and $P(t)$}
 
 For the pendulum equation there is a critical initial velocity at which the coordinate $q(t)$ makes a transition from periodic motion to unbounded motion.  This occurs when the modulus parameter  $l =1$.  This yields the condition 
 \bq
 v_c = \sqrt{ \frac{4 I_4^0 \epsilon}{M_0}} . 
 \eq
 Depending on our choice of parameters,  for small enough $\epsilon$, $v_c$ will be in the non-relativistic regime. 
 We choose $\omega$ to be in the stability region for the unforced problem (see Sec. III).  For $g=1$, 
 $m=1$, $\omega_c = 0.697586$ and   
 choosing $\omega =0.9$, then the width of the solitary wave is $1/(2 \beta) = 1.15$.  If we choose $k=0.1$, then the characteristic wave length $2 \pi/k = 62.8 \gg  1/(2 \beta)$.   From Eqs. (\ref{mzero}) and (\ref{h12}) we have that 
 \bq
 Q =  0.968644 ;  ~~~ H_1 = 0.0625108= I_0 ,
 \eq
 \bq
 M_0 = H_1+ \omega Q = 0.934291 . 
 \eq
 The other constants for this initial condition from Eq. (\ref{i45}) are
 \bq
 I_4^0 = 0.94632;~~ I_5^0= 0.433477 . 
 \eq
 

We have first  compared the analytical solution Eq. (\ref{qccnr}) of the pendulum equation with the numerical solution of Eq. (\ref{eq733}). For $\epsilon < 0.1$ the results are practically identical, for 
 $\epsilon \ge 1$ deviations occur. 
 
 Specifically we have chosen the initial condition $q(0)=0, \dq = v_0$ for the three  cases  (1) $v_0 \ll v_c \ll1 $  and then 
 $v_0$ slightly below (2)  and above (3) the critical value $v_c$, namely
 \bq
 v_0 = v_c \mp .001 . 
 \eq
 Choosing $\epsilon = 0.001$ yields $v_c =0.0636619$ which is in the non-relativistic regime, so we expect Eq. (\ref{qccnr})  to hold. 
 In Fig. \ref{sp1}   we show that for $v_0 = 0.01$ the analytic non-relativistic  result and 
 the numerical solution of  
Eq.\ (\ref{eq733}) give the same results as the solution of the NLDE.  We also see that the shape of the charge density does not change in time. 
In Figs. \ref{sp2} and \ref{sp3} we show that just below and above the critical velocity, respectively, the 
analytical result (\ref{qccnr}) agrees with the numerical solution of Eq.\ (\ref{eq733}), but both results differ 
very slightly from the simulation results. 
 
\begin{figure}[ht!]
\begin{center}
\begin{tabular}{c}
\includegraphics[width=8.0cm]{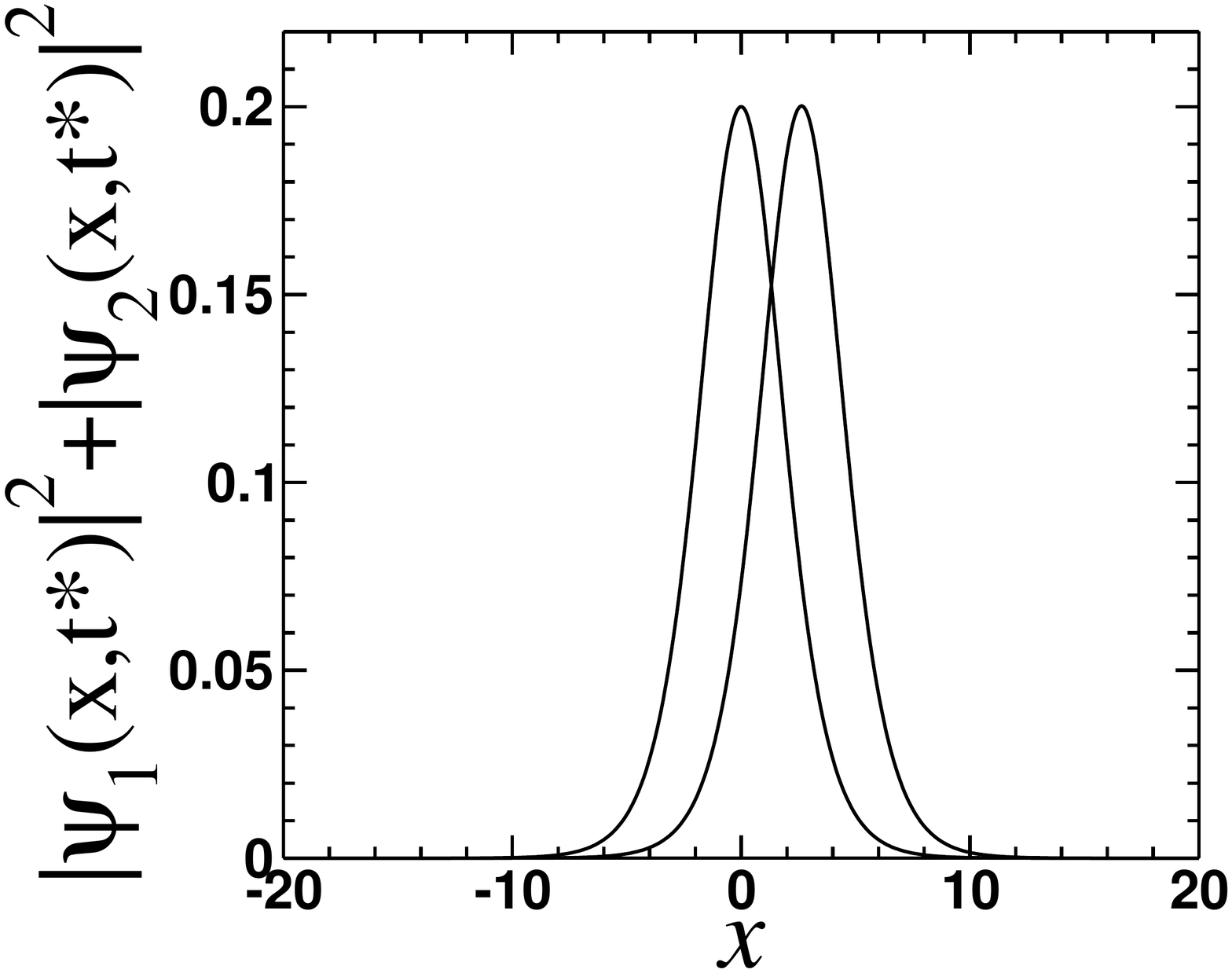}  
\\
\\
\\ 
\includegraphics[width=8.0cm]{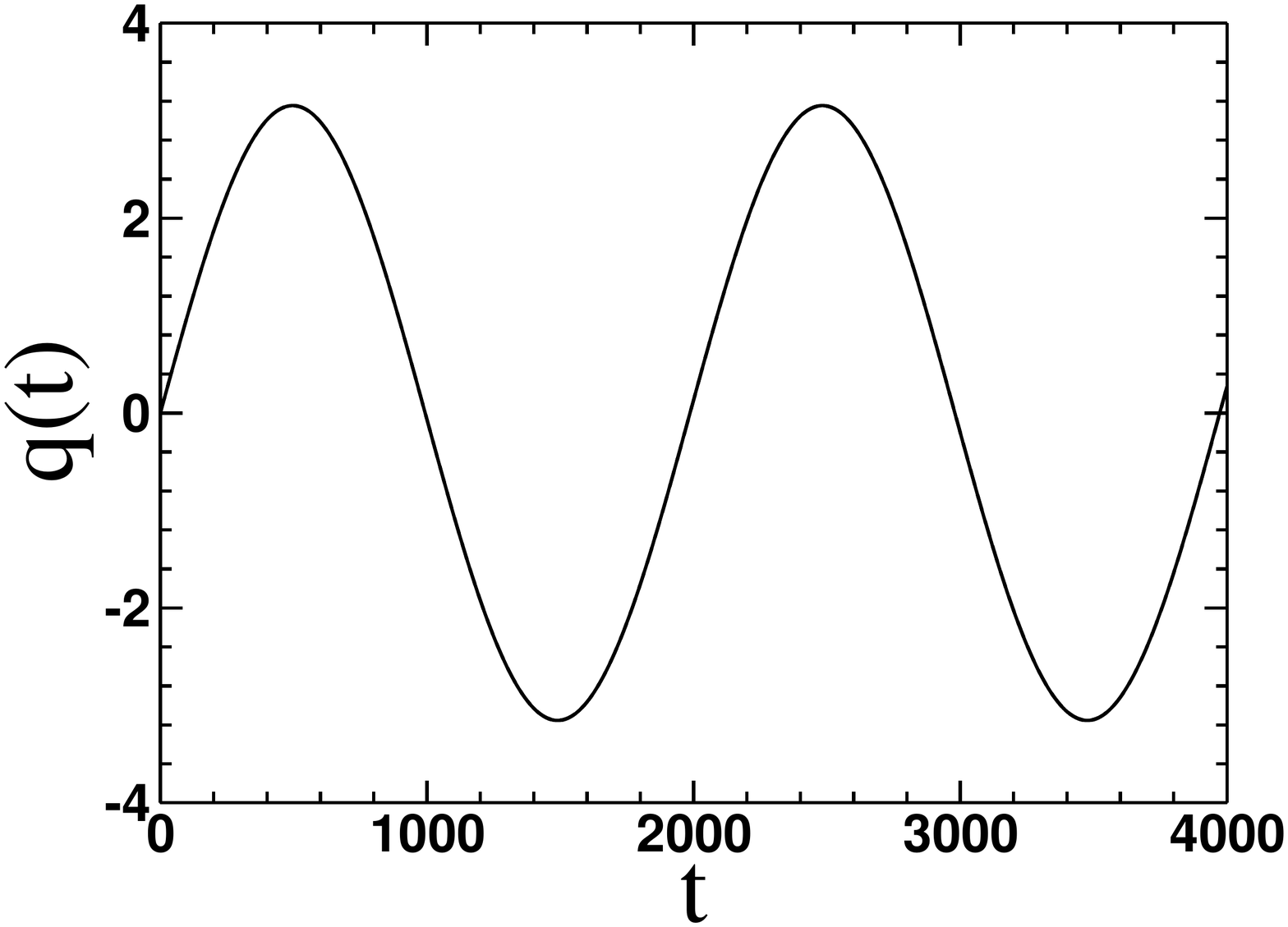}  \\
\\
\\
\includegraphics[width=8.0cm]{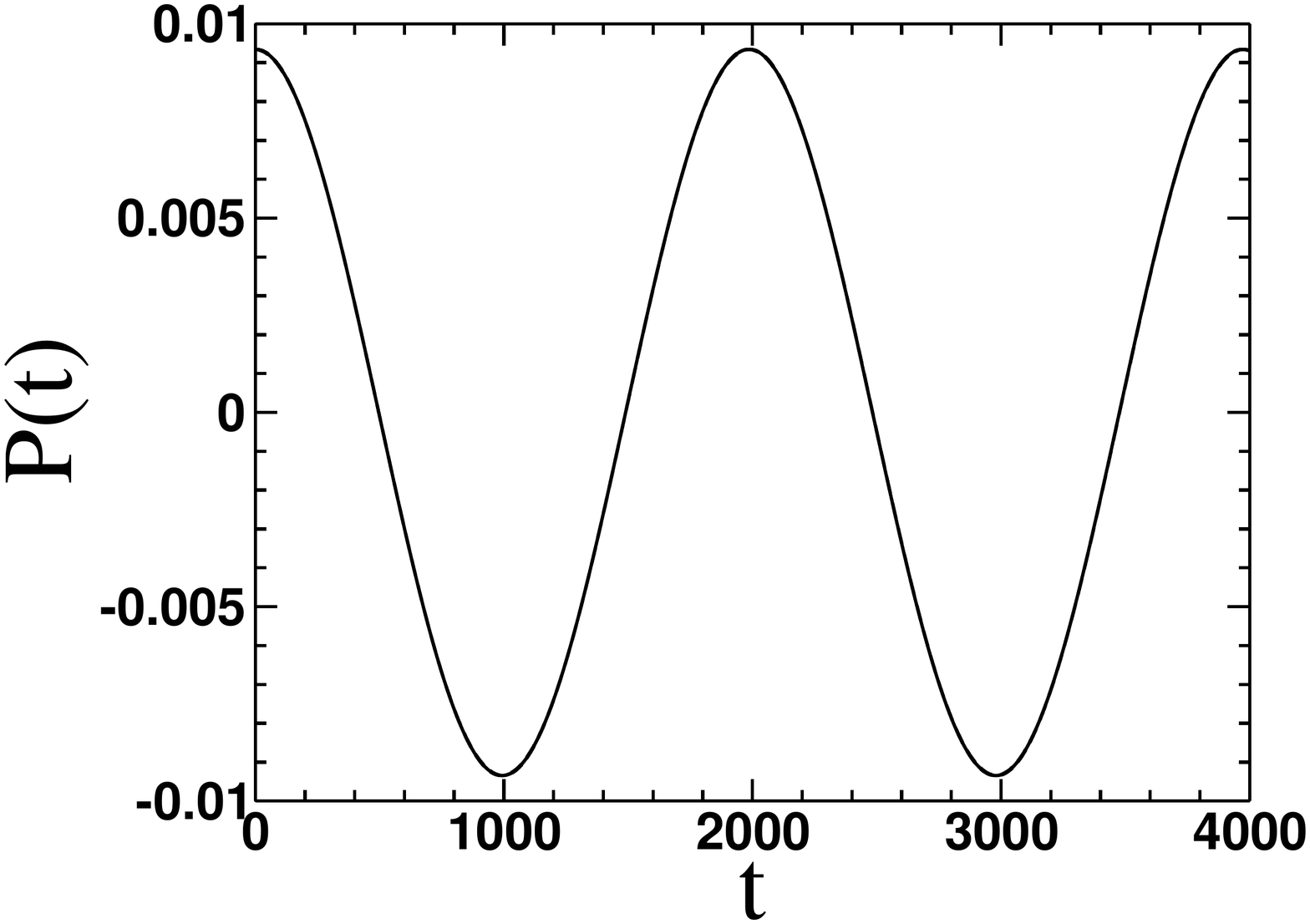}  
\end{tabular}
\end{center}
\caption{ 
Periodic potential, $V(x)=-\epsilon \cos(k x)$, very low initial velocity. Upper panel: Charge density $\rho_Q$ at 
$t^{*}=0;2666.6$.  
Middle panel: solitary wave position $q(t)$, from the numerical solution of Eq. (\ref{eq733}) (solid line), 
approximate analytical expression (\ref{qccnr}) (dotted line),  
and from numerical simulations (dashed line) of the forced 
NLDE. The three curves are super-imposed.   
Lower panel: momentum $P(t)$, from the numerical solutions of Eq. (\ref{eq733}) (solid line)   
and from numerical simulations (dashed line) of the forced 
NLDE. The curves are super-imposed.
Parameters: $g=1$, $m=1$, $\omega=0.9$, $\epsilon=0.001$ and $k=0.1$. 
Initial condition: exact solitary wave of the unperturbed NLDE with 
 initial velocity $v(0)=0.01$.}
\label{sp1} 
\end{figure}

\begin{figure}[ht!]
\begin{center}
\begin{tabular}{c}
\includegraphics[width=8.0cm]{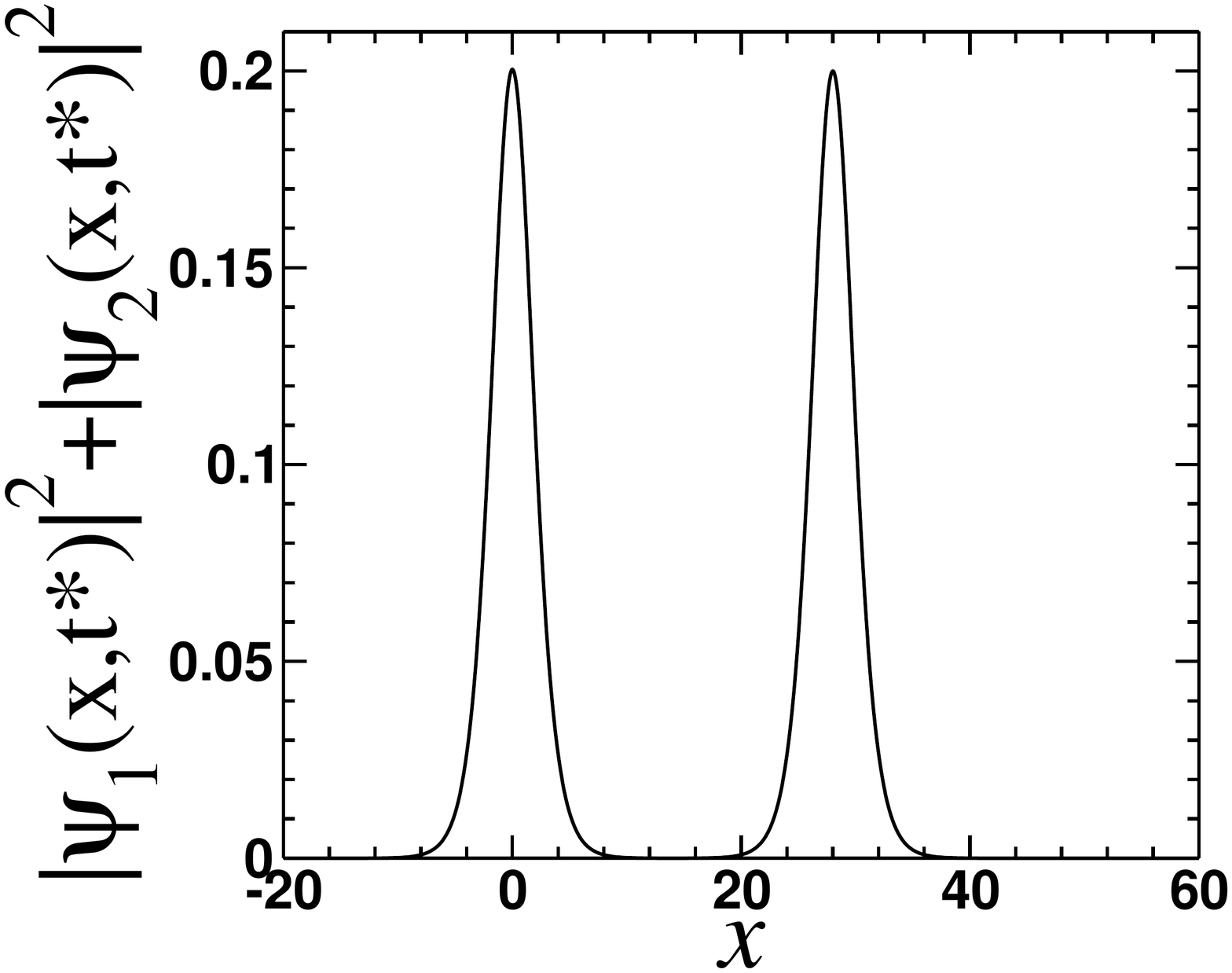}   
\\
\\
\\ 
\includegraphics[width=8.0cm]{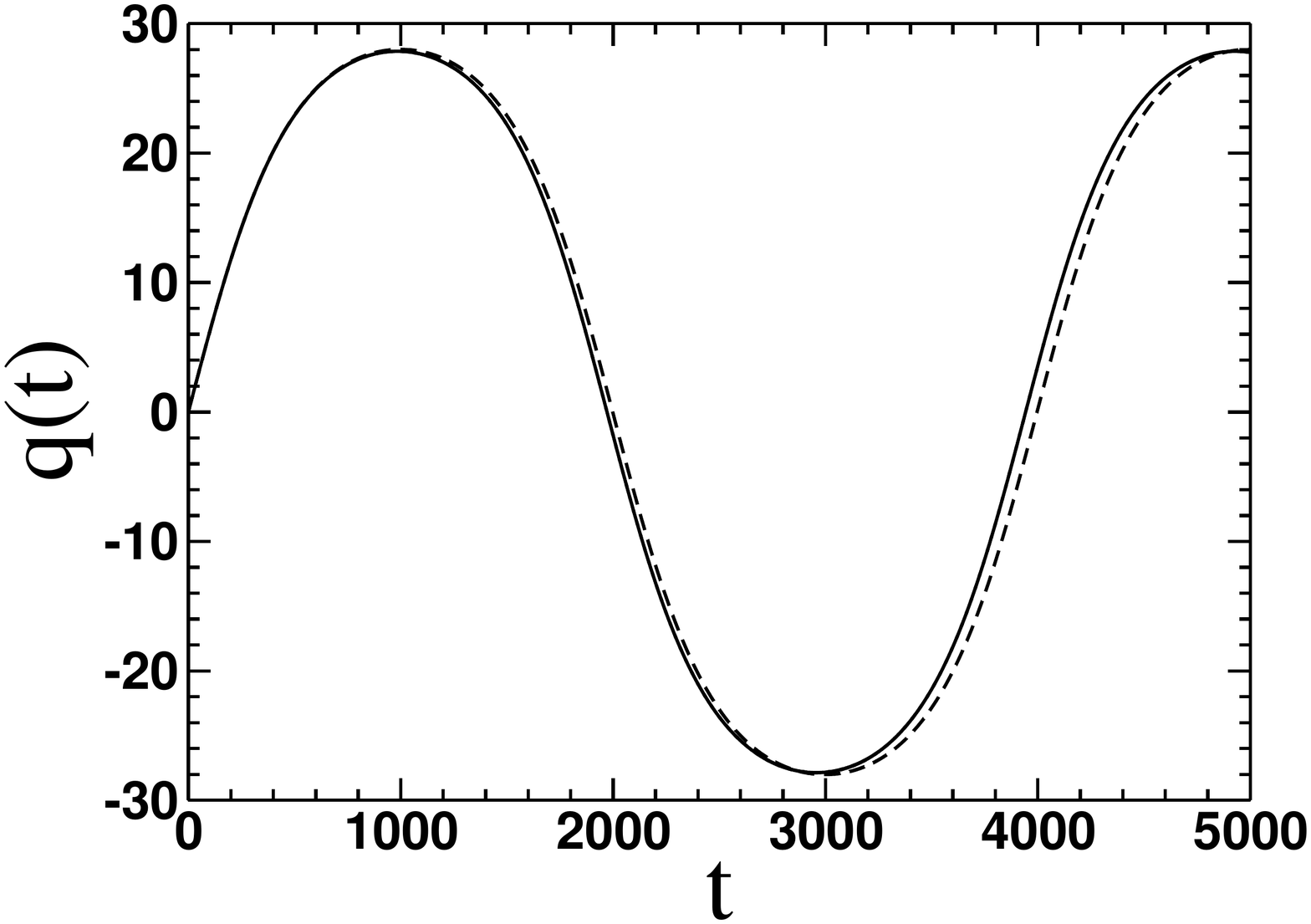}  
\\
\\
\\
\includegraphics[width=8.0cm]{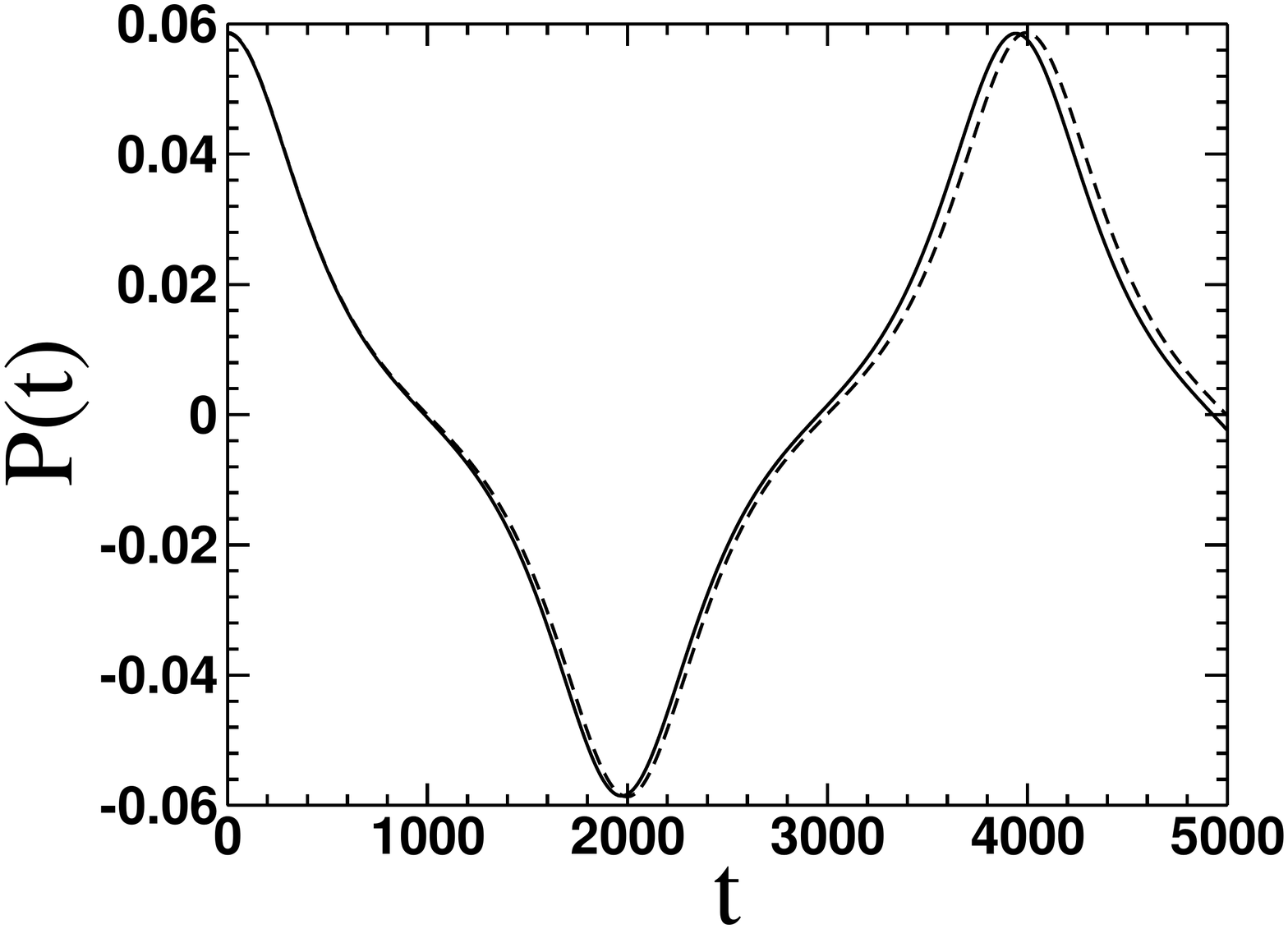}  
\end{tabular}
\end{center}
\caption{Spatially periodic potential, $V(x)=-\epsilon \cos(k x)$, initial velocity just below $v_c$. Upper panel: Charge density $\rho_Q$ at 
$t^{*}=0;5000$. 
Middle panel: solitary wave position $q(t)$, from the numerical solutions of Eq. (\ref{eq733})  (solid line), 
approximate analytical expression (\ref{qccnr}) (dotted line),  
and from numerical simulations (dashed line) of the forced 
NLDE. Solid and dotted lines are super-imposed.   
Lower panel: momentum $P(t)$, from the numerical solution of Eq. (\ref{eq733})  (solid line)   
and from numerical simulations (dashed line) of the forced 
NLDE. 
Parameters: $g=1$, $m=1$, $\omega=0.9$, $\epsilon=0.001$ and $k=0.1$. 
Initial condition: exact solitary wave of the unperturbed NLDE with 
 initial velocity $v(0)=0.0626619$.}
\label{sp2} 
\end{figure}

\begin{figure}[ht!]
\begin{center}
\begin{tabular}{c}
\includegraphics[width=8.0cm]{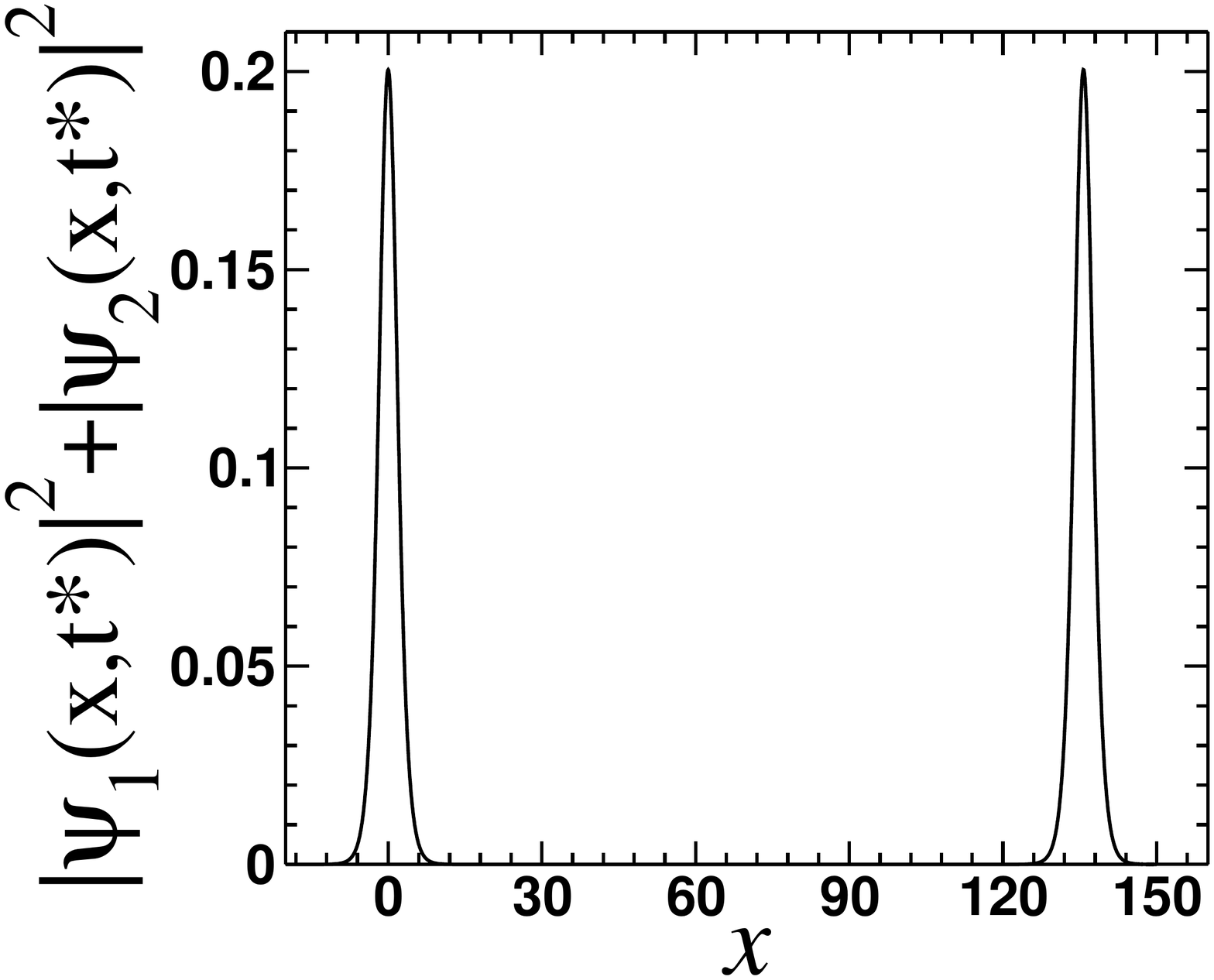}  
\\
\\
\\ 
\includegraphics[width=8.0cm]{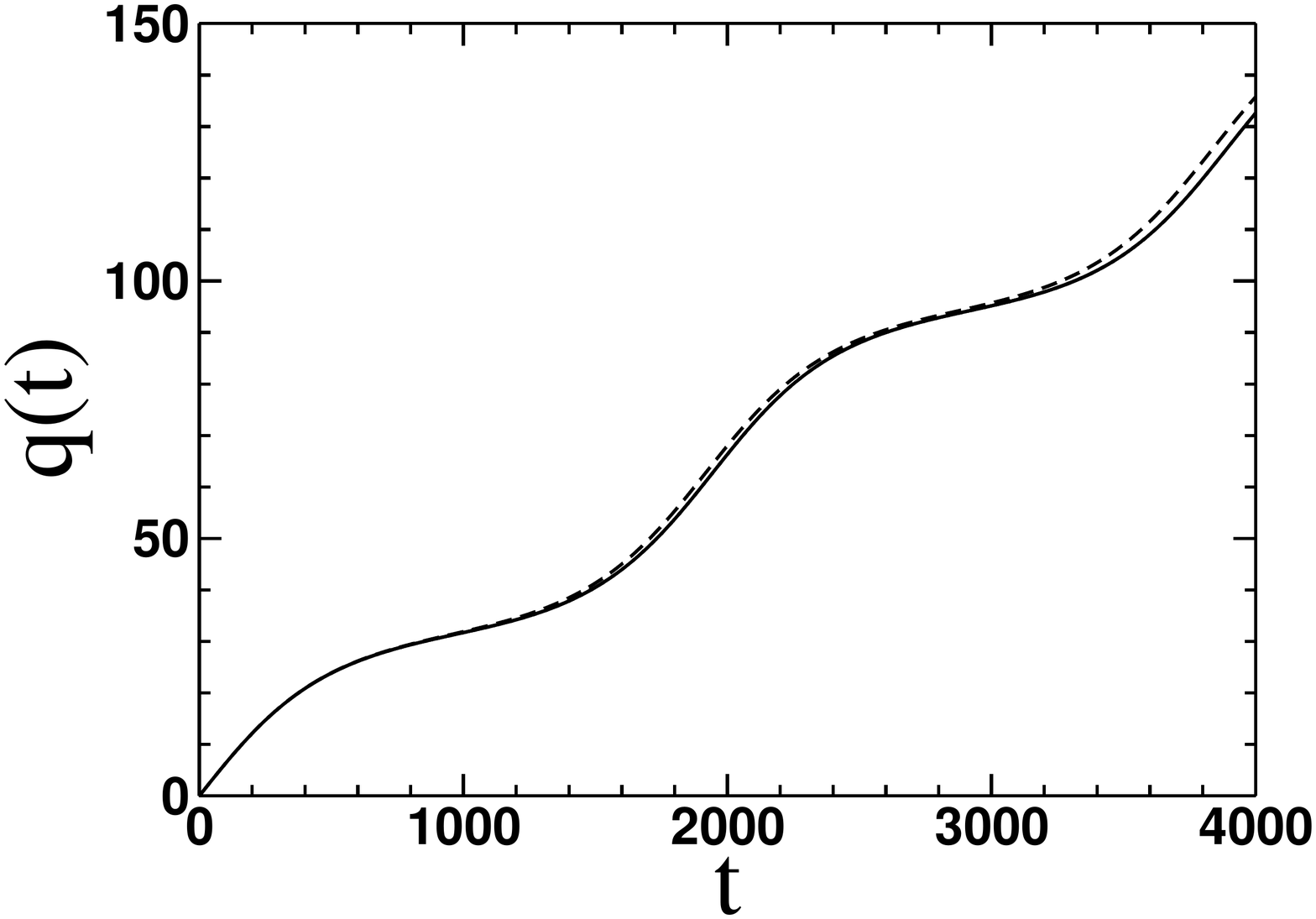}  
\\
\\
\\ 
\includegraphics[width=8.0cm]{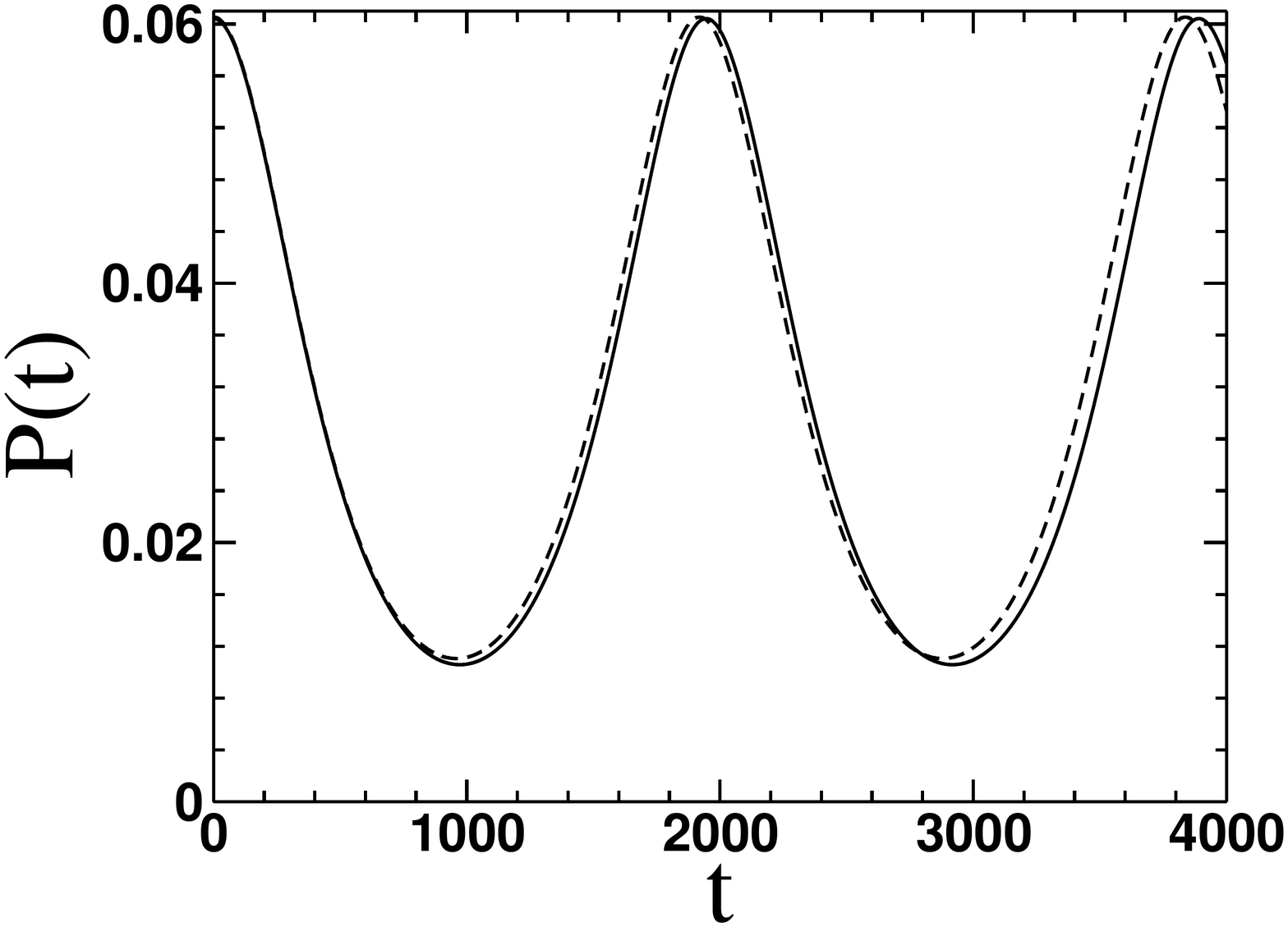}  
\end{tabular}
\end{center}
\caption{Spatially periodic potential, $V(x)=-\epsilon \cos(k x)$, initial velocity just above $v_c$. 
Upper panel: Charge density $\rho_Q$ at 
$t^{*}=0;4000$. 
Middle panel: solitary wave position $q(t)$, from the numerical solutions of Eq. (\ref{eq733}) (solid line), 
approximate analytical expression (\ref{qccnr}) (dotted line),  
and from numerical simulations (dashed line) of the forced 
NLDE. Solid and dotted lines are super-imposed.   
Lower panel: momentum $P(t)$, from the numerical solution of Eq. (\ref{eq733}) (solid line)   
and from numerical simulations (dashed line) of the forced 
NLDE. 
Parameters: $g=1$, $m=1$, $\omega=0.9$, $\epsilon=0.001$ and $k=0.1$. 
Initial condition: exact solitary wave of the unperturbed NLDE with 
 initial velocity $v(0)=0.0646619$.}
\label{sp3} 
\end{figure}
A summary of the result of our simulations of solitary waves in different external fields are found in Table \ref{table}.
\begin{table}
\begin{tabular}{|c|c|c|}
\hline
\hline
Potential & Cases & Results \\
\hline
$V(x)=-V_1x$ & $\omega = 0.9 > \omega_c=0.697586$ & Stable soliton, width Lorentz contracted,\\ 
 & $V_1= 10^{-2}$, $10^{-3}$, $10^{-4}$ & height increases. \\ 
 & $\omega=0.3 < \omega_c$ &  Asymmetric shape, metastable for $t\le 100$, \\ 
 & $V_1=0.01$ & unstable for $t\gtrsim 110$. \\
 & $\omega=0.3 < \omega_c$  & Metastable for $t\le 100$,  splits into \\ 
 & $V_1=0.0001$ & two solitons and radiation for $t\gtrsim 120$. \\
\hline
$V(x)=\frac{1}{2} V_2 x^2$ & $\omega=0.9 > \omega_c$ & Stable soliton, harmonic oscillations.\\ 
 & $V_2=0.0001$, $v_0=0.1$ &  \\ 
 & $\omega=0.3 < \omega_c$, & Metastable for $t \lesssim 350$,  \\  
 &  $V_2=0.0001$,  $v_0=0.9$ &  unstable for $t\gtrsim 350$. \\ 
 & $\omega=0.3<\omega_c$, & Metastable for $t\lesssim 120$,  splits into\\ 
 & $V_2=0.0001$,  $v_0=0.1$ & two solitons and radiation for $t\gtrsim 120$.  \\
\hline
$V(x)=-\epsilon\cos(kx)$ & $\omega=0.9 > \omega_c$, $k=0.1$, $\epsilon=0.0001$ & Stable soliton, harmonic oscillations.\\ 
 & $v_0=0.1\ll v_c=0.0636619$  &  \\ 
 & $\omega=0.9 > \omega_c$, $k=0.1$, $\epsilon=0.0001$ & Stable soliton, very anharmonic \\
 & $v_0=v_c-0.001$ &  oscillations. \\ 
 & $\omega=0.9 > \omega_c$, $k=0.1$, $\epsilon=0.0001$ & Stable soliton, translational motion \\
 & $v_0=v_c+0.001$ & plus oscillations. \\ 
\hline
\hline
\end{tabular}
\caption{Simulation results for three simple potentials using different
parameter sets; $g=1$ and $m=1$.} \label{table} 
\end{table}

\section{Conclusions} 
In this study we have reviewed exact solutions to the NLDE with scalar-scalar interactions of the form $\frac{ g^2}{\kappa+1} ( {\bPsi} \Psi)^{\kappa+1}$  and have used the form of these solutions as variational wave functions for studying the problem with weak external electromagnetic fields. We have  introduced a collective coordinate method for studying the time evolution of these solitary waves in external fields and determined simple equations for the collective coordinates that parallel those of a relativistic point particle. We found that unless (or until)  the solitary waves displayed an instability the collective coordinates describing the position and momentum in the CC equations gave remarkably good agreement with their  counterparts from our simulations. 
 We then  presented a generalization of a dynamical stability criterion, based only on solving the CC equations, that was useful in studying the stability of solitary waves in the forced NLSE problem.  
For our simulations of the exact evolution as well as the evolution of the collective coordinates we concentrated on $\kappa=1$.  For the forcing terms we used simple 
test potentials such as ramp, harmonic and 
periodic potentials.   In many instances we found that the instability of the  solitary wave solution was related to the metastability of the solitary wave in the absence of external forces and the critical time for breakup was quite close to the time found for the unforced  problem.    

We had hoped that a generalization of the method used to map out domains of instability in the NLSE  using the much simpler  solutions of the collective coordinate problem would also work for the NLDE equation.  Unfortunately for the problems we studied we obtained  $ \frac{dp} {d\dot{q}}  >0 $, which fulfills a necessary condition for stability so that
this method did not give any information about instabilities.  What we did  find using the collective coordinate approximation was that starting with exact solutions of the unforced problem, these solitary waves maintained shape in the CC approximation apart from the parameters becoming functions of time. The collective variables in the simulations, namely $q(t)$ and $P(t)$, were smooth functions for a reasonable period of time, even in the case when the solitary waves were only metastable.  When these collective variables become rapidly oscillating and/or diverging from their values found in the collective coordinate calculation, then that  ``defined" the onset of the instability.  The criterion we  used for the onset of instability using the collective coordinates  is a sufficient condition and we did not find any cases where the condition for this dynamic instability was satisfied. Possibly this is a result of the fact that external fields are different from external sources.  For the external source problem, we would expect in the non-relativistic regime that we would recover the known results for the forced NLSE with source terms due to the arguments of Comech  [\onlinecite{comech}].

The simulations in this paper are confined to the $\kappa=1$ case.  The numerical stability of solitary waves in the absence of external potentials for general $\kappa$ will be presented in a subsequent publication \cite{Niurka}.  The semiclassical reduction of 
NLDE to NLSE and implications for solitary wave stability have been recently discussed in a rigorous fashion by Comech [\onlinecite{comech}].  Our numerical findings  \cite{Niurka} agree with his analysis in the non-relativistic regime.  
\section{Acknowledgment} 

This work was supported in part by the U.S. Department of Energy.  F.G.M. acknowledges the 
hospitality of the Mathematical Institute of the University of Seville (IMUS) and of the Theoretical 
Division and Center for Nonlinear Studies at Los Alamos National Laboratory and financial 
support by the Plan Propio of the University of Seville and by Junta de Andalucia.  N.R.Q. 
acknowledges financial support from the Humboldt Foundation through Research Fellowship for Experienced Researchers SPA 1146358 STP and by the MICINN through 
FIS2011-24540, and by Junta de Andalucia under Projects No. FQM207, No. FQM-00481, No. 
P06-FQM-01735, and No. P09-FQM-4643. 

\section*{Appendix: Rest Frame Identities}
In the rest frame, energy-momentum conservation for the solitary wave solutions for the NLDE with the external Vector potential set to zero  leads to identities among the various integrals  that arise concerning the variational wave function variables $A(x), B(x)$. 
The Lagrangian is given by 
 \bq
L =  \frac{i}{2} \left[ \bPsi \gamma^{\mu} \partial_{\mu} - \partial_{\mu}  \bPsi \gamma^{\mu} \Psi 
\right] - m  \Psi 
+ \frac{g^2}{\kappa+1} (\bPsi \Psi)^{\kappa+1}  \>.
\eq
In the rest frame, the wave function is given by 
\bq
\Psi_0 =  \psi e^{-i \omega t}= \left(  \begin{array} {cc}
      A(x) \\
      i ~B(x) \\ 
   \end{array} \right) e^{-i \omega t}.
   \eq
   where $A(x)$ and $B(x)$ are given by  Eq. (\ref{AB}) 
The energy-momentum conservation is given by Eq. (\ref{emc})
and leads to  two independent equations. The first is
 \bq
 \partial _0 T^{01} + \partial_x T^{11} =0.
\eq
In the rest frame $T^{01} $ is independent of time, so that  $T^{11} ={ \rm constant}$.  If the solution goes to zero at infinity then the constant is zero.  We have then the relationship:
\ba
T^{11} &&= \frac{i}{2} [\bPsi \gamma^x \partial^x \Psi - \partial^x \bPsi \gamma^x \Psi ] + L \nonumber \\
&&= \omega \psi^\dag \psi - m \bpsi \psi+  \frac{g^2}{k+1} (\bpsi \psi)^{k+1} 
=0.
\ea
Integrating over space we get the relations:
\bq
\omega Q -m I_1 + \frac{g^2}{\kappa+1} I_2 
 =0  . \label{identity1}
\eq
The second conservation law is 
\bq
 \partial _0 T^{00} + \partial_x T^{x0} =0 , 
\eq
which leads to the conservation of energy.  The energy of the solitary wave in the rest frame defines the rest mass $M_0$
\bq
E = \int T^{00} dx  = M_0 . 
\eq
We have that 
\ba
T^{00} = && - \frac{i}{2} [\bpsi \gamma^x \partial_x \psi - \partial_x \bpsi \gamma^x \psi ] \nonumber \\
&& +m \bpsi \psi -  \frac{g^2}{k+1} (\bpsi \psi)^{k+1}
 \nonumber \\
&& = (A B_x -B A_x) + m(A^2-B^2) -  \frac{g^2}{k+1}(A^2-B^2)^{\kappa+1}  
\ea
Integrating we obtain
\bq
M_0 = I_0 + mI_1 - \frac{g^2}{k+1} I_2
\eq
Using the identity of Eq. (\ref{identity1}), we then have 
\bq
M_0 = I_0 + \omega Q . 
\eq

\end{document}